\documentclass[aps,prb,reprint,longbibliography,superscriptaddress,nofootinbib,floatfix]{revtex4-2}
\usepackage{amsmath,amssymb,mathrsfs,graphicx,booktabs}
\usepackage{float}

\usepackage[colorlinks,citecolor=blue,urlcolor=blue,linkcolor=blue]{hyperref}
\hypersetup{pdftitle={Orbital cancellation and reentrant localization in a quasiperiodic trimer ladder},pdfauthor={Xin Wang, Pei-Jie Chang, Quanfeng Lu, Muxi Zheng, Dong Ruan, Gui-Lu Long}}
\graphicspath{{figures/}}
\newcommand{\MainCells}{1597}
\newcommand{\CheckCells}{987}
\newcommand{\MaxCells}{100000}
\newcommand{\NumSizes}{16}

\newcommand{\DynMaxTime}{10^5}

\newcommand{\ScalingScanPoints}{281}
\newcommand{\FirstWindowStart}{4.1800}
\newcommand{\FirstWindowEnd}{5.0700}
\newcommand{\FirstWindowWidth}{0.8900}
\newcommand{\SecondWindowStart}{11.7200}
\newcommand{\SecondWindowEnd}{12.3650}
\newcommand{\SecondWindowWidth}{0.6450}

\newcommand{\Icell}{\mathcal I}
\newcommand{\Ncell}{\mathcal N}
\makeatletter
\let\check@deferlist@stuck@original\check@deferlist@stuck
\def\check@deferlist@stuck#1{%
  \ifx#1\do@startpage
    \begingroup
      \@tempswafalse
      \def\@elt##1{\@if@notdblfloat{##1}{}{\@tempswatrue}}%
      \@deferlist
      \if@tempswa
        \endgroup
        \check@deferlist@stuck@original#1%
      \else
        \endgroup
      \fi
  \else
    \check@deferlist@stuck@original#1%
  \fi
}
\makeatother
\begin{document}
\title{Orbital cancellation and reentrant localization in a quasiperiodic trimer ladder}

\author{Xin Wang}
\thanks{Xin Wang and Pei-Jie Chang contributed equally to this work.}
\affiliation{School of Physical Science and Technology, Nantong University, Nantong, 226019, China}

\author{Pei-Jie Chang}
\email{cpj22@mails.tsinghua.edu.cn}
\affiliation{State Key Laboratory of Low-Dimensional Quantum Physics and Department of Physics, Tsinghua University, Beijing 100084, China}

\author{Quanfeng Lu}
\affiliation{State Key Laboratory of Low-Dimensional Quantum Physics and Department of Physics, Tsinghua University, Beijing 100084, China}

\author{Muxi Zheng}
\affiliation{State Key Laboratory of Low-Dimensional Quantum Physics and Department of Physics, Tsinghua University, Beijing 100084, China}

\author{Dong Ruan}
    \affiliation{State Key Laboratory of Low-Dimensional Quantum Physics and Department of Physics, Tsinghua University, Beijing 100084, China}
    \affiliation{Frontier Science Center for Quantum Information, Tsinghua University, Beijing 100084, China}

\author{Gui-Lu Long}
    \email{gllong@tsinghua.edu.cn}
    \affiliation{State Key Laboratory of Low-Dimensional Quantum Physics and Department of Physics, Tsinghua University, Beijing 100084, China}
    \affiliation{Frontier Science Center for Quantum Information, Tsinghua University, Beijing 100084, China}
    \affiliation{Beijing Academy of Quantum Information Sciences, Beijing 100193, China}
    \affiliation{Beijing National Research Center for Information Science and Technology, Beijing 100084, China}

\begin{abstract}
Uniform orbital detuning can restore spatial extent even as quasiperiodic modulation strengthens. In a trimer ladder with fixed hoppings and orbital modulation signs, increasing detuning and modulation amplitude together drives successive avoided crossings that redistribute orbital weights, suppressing the projected onsite potential in both lower molecular bands while leaving the third strongly modulated. Complete-band participation scaling supports two reentrant high-participation windows, with wave-packet propagation corroborating spatial recovery at representative controls over the studied sizes and times. These diagonal cancellations leave residual interband modulation, which generates a self-energy whose magnitude depends on the molecular level separation. Gaussian quadrature improves convergence of closed-walk determinant estimates used to cross-check energy-resolved attenuation. Comparing the coupled and decoupled ladders under the same finite-size participation criterion shows that restoring the full off-diagonal modulation narrows both sampled windows, with a larger absolute contraction of the second.
\end{abstract}
\maketitle
\raggedbottom
\setlength{\textfloatsep}{12pt plus 2pt minus 2pt}
\setlength{\dbltextfloatsep}{12pt plus 2pt minus 2pt}
\setlength{\abovedisplayskip}{8pt plus 2pt minus 4pt}
\setlength{\belowdisplayskip}{8pt plus 2pt minus 4pt}
\setlength{\abovedisplayshortskip}{8pt plus 2pt minus 4pt}
\setlength{\belowdisplayshortskip}{8pt plus 2pt minus 4pt}
\makeatletter
\def\@textbottom{\vskip0pt plus4pt}
\makeatother
\section{Introduction}
\label{sec:intro}
Anderson localization arises from quantum interference in disordered systems~\cite{Anderson1958,Thouless1974,Abrahams1979,LeeRamakrishnan1985,EversMirlin2008}. Quasiperiodic lattices provide a deterministic setting for localization, with incommensurate spatial modulation replacing random disorder. The Harper and Aubry--Andr\'e (AA) models relate localization directly to potential strength through a self-dual threshold~\cite{Harper1955,AubryAndre1980}. Generalized potentials, hopping, and continuum structure permit energy-dependent thresholds~\cite{DasSarma1988,Ganeshan2015,Biddle2009,Biddle2010,Boers2007,Li2017,Yao2019,Wang2020mosaic}. For nearest-neighbor hopping, deformed onsite potentials support an exact mobility edge protected by generalized duality~\cite{Ganeshan2015}. In mosaic lattices, quasiperiodic modulation on equally spaced sites also produces analytically determined mobility edges~\cite{Wang2020mosaic}. Multicomponent models can contain extended, localized, and critical spectral sectors~\cite{Hiramoto1989,Wang2016,Wang2020raman,Li2020,Wang2022Coexist,Zhou2023CriticalME,Zhou2026Unified}.

Quasiperiodic lattices can also respond nonmonotonically to an increasing control parameter, exhibiting reentrant localization when spatial extent recovers after localization has developed. This behavior has been reported in staggered dimerized chains~\cite{Roy2021,Roy2022} and has been observed in photonic quasicrystals~\cite{Vaidya2023}. Reported mechanisms involve composite potentials~\cite{Padhan2022,Qi2023Staggered,Wei2025,ChangAK2025}, relative phases~\cite{LiPhase2023}, longer-range hopping~\cite{Wang2023longrange,Chang2025njp}, and molecular bands~\cite{Tabanelli2024}. Interpolation from the AA model to the Fibonacci chain produces a cascade of band-selective localization and delocalization transitions. Theory and cavity-polariton experiments link this cascade to the emergence of critical wave functions in the Fibonacci limit~\cite{Goblot2020}. Dipolar coupling~\cite{AllardWeick2026Dipolar}, correlated hopping~\cite{Karmakar2025CorrelatedHopping}, and internal or geometric tuning~\cite{Guan2023,Ganguly2025,Shao2026LadderShift} provide further examples. A double-stranded helix exhibits two reentrant regions, with the second described as quasi-extended~\cite{Ganguly2024}.

Among these mechanisms, internal orbital mixing changes the modulation experienced by each effective channel through its orbital composition. In symmetric ladders, a rotation of the internal basis reduces the Hamiltonian to AA channels~\cite{Maiti2017,WangYangSong2021Ladder}; rung modulation~\cite{Goswami2026Ladder} and dot decoration~\cite{PalChakrabarti2014} also permit delocalization. Opposite orbital modulation signs can cancel the projected quasiperiodic onsite potential within each channel. In flat-band models, elimination of the remaining coupled channel generates a squared potential~\cite{Bodyfelt2014,Danieli2015}, and competing harmonics can produce reentrance~\cite{LuXu2024CrossStitch}. Coupling a chain with extended states to one with localized states can produce critical states in their overlapping spectra. Multifractal diagnostics resolve this regime across several choices of interchain coupling and quasiperiodic potential~\cite{Lin2023CoupledChains}. The spatial response therefore depends on both the diagonal projection and the residual interchannel coupling.

The coupled-channel response can be tuned through internal-state parameters, as in spin-orbit-coupled quasiperiodic lattices~\cite{KohmotoTobe2008,Zhou2013SpinOrbit}. In a Raman lattice, a uniform Zeeman term drives reentrant delocalization by shifting mobility edges and splitting the localized region~\cite{Wei2026RamanReentrance}. In a two-leg ladder at weak quasiperiodic modulation, staggered detuning on one leg can localize the neighboring leg through interleg hybridization. Increasing the detuning further produces reentrant spreading at fixed modulation amplitude~\cite{Giri2026Detuning}.

The Zeeman and leg-selective detuning results above concern fixed quasiperiodic amplitude, while the symmetric-ladder and flat-band reductions use exact channel separation or a nondispersive eliminated channel. When detuning and quasiperiodic amplitude increase together, orbital redistribution can suppress the diagonal projection even as the microscopic modulation grows. Residual coupling between dispersive molecular channels then competes with this suppression. The resulting problem is whether the projected spatial recoveries survive in the coupled ladder and how residual scattering modifies the corresponding windows.

A trimer ladder with opposite modulation signs on the central and outer orbitals permits two such cancellations along a joint detuning--amplitude path, with all hoppings fixed. The path labels a family of static Hamiltonians, with dynamics evaluated at fixed parameters. Molecular projection identifies the candidate windows, and complete-band participation scaling and propagation test their survival in the coupled ladder. Closed-walk determinant calculations, evaluated by Gaussian quadrature to improve convergence, provide independent checks of energy-resolved attenuation. A comparison of coupled and decoupled models under identical participation criteria isolates the changes caused by residual interband modulation. The self-energy analysis relates the effective Hamiltonian correction to molecular level separation, while independent detuning--amplitude paths examine the persistence of the two high-participation regions.

The remainder of this paper is organized as follows. Section~\ref{sec:model} introduces the trimer ladder and its isolated-channel predictions. Section~\ref{sec:spectrum} presents participation scaling and dynamics of the full ladder. Section~\ref{sec:interband} develops the interband self-energy and compares coupled and decoupled participation windows alongside energy-resolved attenuation. Section~\ref{sec:scope} examines the control plane and nearby monotone paths. Section~\ref{sec:conclusion} summarizes the results. The appendices give the determinant formulation, derivations, numerical protocols, and convergence checks.

\section{Model and orbital cancellation}
\label{sec:model}
\label{sec:orbital}
\subsection{Ladder Hamiltonian and control path}
The noninteracting trimer ladder has the Hamiltonian
\begin{align}
H(u,v)={}&J\sum_{m=0}^{L-2}\left(\boldsymbol c_{m+1}^{\dagger}\boldsymbol c_m+\mathrm{H.c.}\right)\nonumber\\
&+J\sum_{m=0}^{L-1}\boldsymbol c_m^\dagger
\left[\Omega K(u)+v\cos x_m\,Q\right]\boldsymbol c_m,
\label{eq:H}
\end{align}
where $\boldsymbol c_m=(a_m,b_m,c_m)^T$ collects the annihilation operators for orbitals $A$, $B$, and $C$ in longitudinal cell $m=0,\ldots,L-1$. The longitudinal hopping amplitude is $J$, while $u$ controls the uniform orbital detuning and $v$ sets the quasiperiodic modulation amplitude. The intracell matrix and orbital modulation are specified by
\begin{align}
K(u)={}&\begin{pmatrix}u&g_1&0\\g_1&b&g_2\\0&g_2&c-u\end{pmatrix},
\qquad Q=\operatorname{diag}(1,-1,1).\label{eq:K}
\end{align}

\begin{figure}[!t]
\includegraphics[width=\columnwidth]{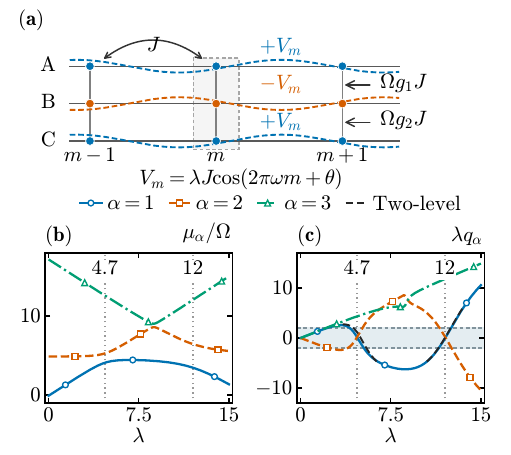}
\caption{(a) Trimer ladder with longitudinal hopping $J$, intracell hoppings $\Omega g_1J$ ($A$--$B$) and $\Omega g_2J$ ($B$--$C$), and onsite modulation $\pm V_m$. The uniform detunings and reference parameters are given in Eqs.~\eqref{eq:K} and \eqref{eq:pars}. (b) Molecular eigenvalues $\mu_\alpha/\Omega$. (c) Projected modulation amplitudes $\lambda q_\alpha$ [Eq.~\eqref{eq:charge}]. Black dashed segments show the local two-level approximations $\lambda q_-^{AB}$ and $\lambda q_-^{BC}$ [Eq.~\eqref{eq:twocross}]. Horizontal gray dashed lines bound the shaded isolated-channel subcritical strip $|\lambda q_\alpha|<2$. Colors and symbols identify the same molecular bands in (b,c); vertical dotted lines mark the representative samples $\lambda=4.7,12$.}
\label{fig:model}
\end{figure}

The modulation phase is $x_m=2\pi\omega m+\theta$, where $\omega$ is the quasiperiodic frequency and $\theta$ is the phase offset; $\Omega$ sets the molecular energy scale. The reference path is $u=v=\lambda$, with $H(\lambda)=H(\lambda,\lambda)$ and $K=K(\lambda)$. Open boundary conditions and units $J=\hbar=1$ are used, with unit longitudinal cell spacing. The chain contains $L$ cells and $N=3L$ orbitals. Energies, including $E$ and the molecular eigenvalues $\mu_\alpha$ defined below, are expressed in units of $J$; $Jt$ is dimensionless time. All parameters inside the brackets in Eq.~\eqref{eq:H} are dimensionless, and $\lambda\geq0$.

The reference parameters are
\begin{equation}
\begin{gathered}
\Omega=100,\quad g_1=0.6,\quad g_2=1.5,\\
b=5,\quad c=17,\quad \omega=\frac{\sqrt5-1}{2}.
\end{gathered}
\label{eq:pars}
\end{equation}

As shown in Fig.~\ref{fig:model}(a), longitudinal hopping connects the same orbital in neighboring cells, while intracell hopping couples $A$ to $B$ and $B$ to $C$.

Along the reference path $u=v=\lambda$, increasing $\lambda$ strengthens the quasiperiodic modulation while raising the uniform onsite energy of orbital $A$ and lowering that of orbital $C$. The uniform onsite energy of orbital $B$ and all hopping amplitudes remain fixed.

\subsection{Molecular levels and orbital weights}
\label{sec:mechanism}
At fixed $\lambda$, a cell-independent orthogonal matrix $R$ diagonalizes the uniform $3\times3$ intracell, or molecular, Hamiltonian:
\begin{equation}
R^T\Omega K R=\operatorname{diag}(\mu_1,\mu_2,\mu_3).
\end{equation}
The eigenvalues are ordered as $\mu_1\leq\mu_2\leq\mu_3$.
At fixed $u$ and $v=0$, each molecular level generates a band of $L$ chain states.
The index $\alpha=1,2,3$ also labels the well-separated bands of the full model considered below, although the modulation generally couples these molecular channels.
Because the same $R$ is used in every cell and the longitudinal hopping is proportional to the identity, neighboring molecular states with the same $\alpha$ remain connected by hopping $J$. The rotation transforms the modulation matrix to $\widetilde Q=R^T Q R$, whose diagonal entries are
\begin{align}
q_\alpha=\widetilde Q_{\alpha\alpha}
&=|R_{A\alpha}|^2-|R_{B\alpha}|^2+|R_{C\alpha}|^2\nonumber\\
&=1-2|R_{B\alpha}|^2.
\label{eq:charge}
\end{align}

The coefficient $q_\alpha$ is the probability-weighted sum of the orbital modulation signs. Diagonal cancellation occurs at $|R_{B\alpha}|^2=1/2$, where the central-orbital weight equals the combined outer-orbital weight. The projected quasiperiodic onsite potential $\lambda q_\alpha\cos x_m$ then vanishes in every cell, although the microscopic amplitude $\lambda$ increases along the control path. The off-diagonal entries of $\widetilde Q$ retain interchannel coupling.

The molecular energies in Fig.~\ref{fig:model}(b) show how these weights change along the reference path. Without intracell mixing, the lowest orbital would switch from $A$ to $B$ at $\lambda=b$ and from $B$ to $C$ at $\lambda=c-b$. The finite couplings $g_1$ and $g_2$ turn these crossings into avoided crossings. As $\lambda$ increases, the lowest molecular eigenstate changes continuously from predominantly $A$ to predominantly $B$ and then to predominantly $C$. Its $B$-orbital weight crosses $1/2$ twice, giving two zeros of $q_1$. The second-band projection is also small near these crossings. Direct molecular diagonalization gives the band-1 zeros $4.74382$ and $12.0346$, and the band-2 zeros $4.65449$ and $12.0807$. The representative values $4.7$ and $12$ sample the neighborhoods of these paired zeros. Orthogonality gives
\begin{equation}
\sum_{\alpha=1}^3q_\alpha=3-2\sum_{\alpha=1}^3|R_{B\alpha}|^2=1.
\label{eq:chargesum}
\end{equation}
When both lower-band projections are small, Eq.~\eqref{eq:chargesum} gives $q_3=1-q_1-q_2\simeq1$, leaving the third channel strongly modulated near the paired cancellations.

The intermediate avoided crossing between the two upper molecular levels mainly exchanges $A$ and $C$ weight. Both orbitals carry the same modulation sign, so this exchange does not produce an additional diagonal cancellation.

\subsection{Isolated-channel picture of the two windows}
Neglecting the off-diagonal elements of $\widetilde Q$ gives an AA equation for the amplitude $\phi_m$ in each molecular channel:
\begin{equation}
\phi_{m+1}+\phi_{m-1}+\lambda q_\alpha\cos x_m\,\phi_m
=(E-\mu_\alpha)\phi_m.
\label{eq:effective}
\end{equation}
For energies in the spectrum of this projected equation, the exact AA Lyapunov exponent (LE) per cell~\cite{AubryAndre1980,BourgainJitomirskaya2002LE} is
\begin{equation}
\gamma_\alpha^{(0)}=\max\left\{0,\log\frac{|\lambda q_\alpha|}{2}\right\}.
\label{eq:LEapprox}
\end{equation}
The shaded strip in Fig.~\ref{fig:model}(c), bounded by the gray dashed lines $|\lambda q_\alpha|=2$, marks the isolated-channel subcritical regime from Eq.~\eqref{eq:LEapprox}. The uniform energy $\mu_\alpha$ shifts the channel spectrum without changing this threshold. Both lower-channel amplitudes enter this strip twice as their orbital weights pass through the cancellations.

The two-level approximations near the $A$--$B$ and $B$--$C$ avoided crossings are, respectively,
\begin{equation}
\begin{gathered}
q_-^{AB}\simeq\frac{b-\lambda}{\sqrt{(b-\lambda)^2+4g_1^2}},\\
q_-^{BC}\simeq\frac{\lambda-(c-b)}{\sqrt{[\lambda-(c-b)]^2+4g_2^2}}.
\end{gathered}
\label{eq:twocross}
\end{equation}
Equation~\eqref{eq:twocross} reproduces the sign reversals and local variation of the lower-band projected amplitude obtained from the full $3\times3$ molecular diagonalization [Fig.~\ref{fig:model}(c)]. The differences, including shifts of the cancellation points, arise from coupling to the third orbital omitted in the $2\times2$ reduction. Replacing the prefactor $\lambda$ in $|\lambda q_-|<2$ by its value $\lambda_c>2$ at the crossing gives the isolated-channel window-width estimate $\Delta\lambda\simeq8g/\sqrt{\lambda_c^2-4}$, where $g=g_1$ or $g_2$ for the corresponding crossing. Section~\ref{sec:bandparticipation} compares the overlap of the two lower projected intervals with the joint full-model participation criterion. Residual interband modulation is omitted in this prediction and is analyzed in Sec.~\ref{sec:interband}.

\section{Reentrant behavior of the full ladder}
\label{sec:phenomenology}
\label{sec:spectrum}
\begin{figure}[!htbp]
\includegraphics[width=\columnwidth]{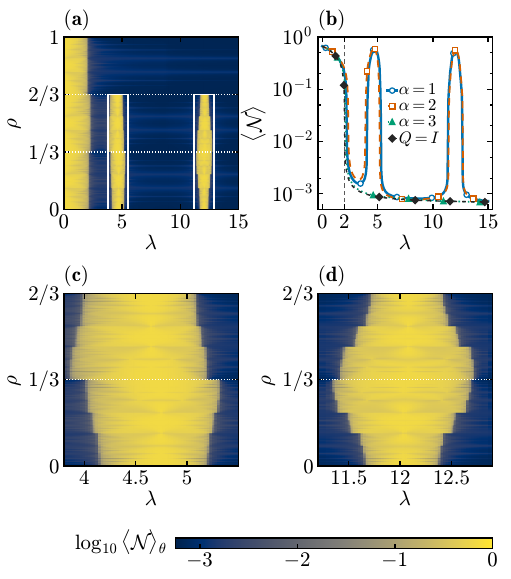}
\caption{Cell NPR at $L=\MainCells$, averaged over four phases. (a) $\log_{10}\langle\Ncell\rangle_\theta$, with the cell NPR averaged over phases at fixed ordered spectral index $\rho=(n-1)/(3L-1)$ before taking the logarithm. (b) Arithmetic means over all $L$ states per band and four phases, and the matched $Q=I$ control, on a logarithmic ordinate; symbols distinguish curves, and the vertical dashed line marks $\lambda=2$. (c,d) Enlargements of the boxed lower-band windows. Panels (a,c,d) share the color scale $[-3.25,0]$. The unsmoothed 281-point scan has spacings from $0.15$ down to $0.005$ near the crossovers; additional path samples refine the NPR-interior endpoints.}
\label{fig:phase}
\end{figure}

The full-ladder spectrum in Fig.~\ref{fig:phase}(a) consists of three separated bands of $L$ states each. The index $n$ labels a full-chain eigenstate, and $\alpha$ labels its energy-ordered band.

For a normalized eigenstate $\psi_{m,s}^{(n)}$, the cell probability, cell inverse participation ratio (IPR), and cell normalized participation ratio (NPR) are
\begin{equation}
p_m^{(n)}=\sum_{s=A,B,C}|\psi_{m,s}^{(n)}|^2,
\quad \Icell_n=\sum_m[p_m^{(n)}]^2,
\quad \Ncell_n=\frac{1}{L\Icell_n}.
\label{eq:ipr}
\end{equation}
These participation ratios are invariant under unitary rotations among the orbitals within each cell. For probabilities uniform over $M$ cells, $\Icell=1/M$ and $\Ncell=M/L$, so cell NPR measures the effective fraction of the chain occupied by a state. Longitudinally extended states have $\Icell\propto L^{-1}$, whereas localized states have a nonzero limiting $\Icell$.

In Fig.~\ref{fig:phase}(a), the phase-averaged cell NPR is plotted against the normalized spectral rank $\rho=(n-1)/(3L-1)$ at $L=\MainCells$. Both lower bands recover high participation near $\lambda=4.7$ and $12$, separated by low-participation regions. The enlargements in panels (c,d) show that both recoveries involve broad portions of the lower bands, while the third band remains at low participation near these parameters. The complete-band averages in panel (b) exhibit the same two peaks.

Replacing $Q$ with the $3\times3$ identity $I$ while retaining the detuning gives an exact control: $\widetilde Q=I$ for every $\lambda$, and the ladder decomposes into three AA chains with amplitude $\lambda$ and uniform energy shifts. All three chains have their transition at $\lambda=2$ [Fig.~\ref{fig:phase}(b)]. Opposite orbital modulation signs enable the vanishing diagonal projections in Eq.~\eqref{eq:charge}.

The two lower-band NPR peaks in Fig.~\ref{fig:phase}(b) occur in the two parameter regions where $\lambda q_1$ and $\lambda q_2$ both enter the shaded strip $|\lambda q_\alpha|<2$ in Fig.~\ref{fig:model}(c). Near both peaks, $\lambda q_3$ remains outside this strip, matching the low NPR of the third band.

\subsection{Participation over complete molecular bands}
\label{sec:bandparticipation}
Finite-size participation scaling distinguishes spatial confinement from extended and critical wave functions~\cite{EdwardsThouless1972,Hiramoto1989,EversMirlin2008}. The geometric mean of $\Icell$ over all states in band $\alpha$ and four uniformly spaced phases is fitted to $\Icell_{\mathrm{geom},\alpha}\propto L^{-D_{2,\alpha}}$, using $L=28657,46368,75025,100000$. The resulting band-averaged exponent can reflect different size dependences within a band as well as finite-size crossover.

The two high-participation windows are identified by requiring at least $90\%$ of eigenstate samples in each lower band, pooled over four phases, to satisfy $\Ncell>0.1$ at both $L=987$ and $1597$. These lengths define the common window criterion used in Table~\ref{tab:matchedendpoints}, while the full sequence of 16 lengths up to $L=100000$ tests participation scaling at representative controls. This criterion gives the sampled intervals $\FirstWindowStart\leq\lambda\leq\FirstWindowEnd$ and $\SecondWindowStart\leq\lambda\leq\SecondWindowEnd$, with widths $\FirstWindowWidth$ and $\SecondWindowWidth$.

Both sampled intervals lie within the overlaps of the two lower channels' subcritical intervals, $4.08800<\lambda<5.13548$ and $11.5637<\lambda<12.5195$. The reduction in sampled width relative to the projected interval is larger for the second window. The full-model endpoints follow from the finite-size NPR threshold, whereas the projected endpoints follow from $|\lambda q_\alpha|=2$. Section~\ref{sec:couplingsensitivity} compares the coupled and decoupled models using the same finite-size NPR criterion.

Six representative parameter values sample the high-participation regimes and localized controls:
\begin{equation}
\lambda=1,\ 3.1,\ 4.7,\ 8,\ 12,\ 14.
\label{eq:points}
\end{equation}

\begin{table}[tp]
\caption{Complete-band participation exponents fitted to $L=28657,46368,75025,100000$, using the geometric mean cell IPR over all band states and four phases. Exponents are rounded to six significant digits. The small excesses above unity accompany a weak decrease of $L\Icell_{\mathrm{geom},\alpha}$ over the fitted lengths. Mean NPR ranges for the two sample groups are reported in Appendix~\ref{app:staticdetails}.}
\label{tab:static}
\begin{ruledtabular}
\normalsize
\begin{tabular}{cccc}
$\lambda$ & $D_{2,1}$ & $D_{2,2}$ & $D_{2,3}$\\\hline
1 & $1.00082$ & $1.00077$ & $1.00084$ \\
3.1 & $7.86374\times10^{-5}$ & $1.18346\times10^{-4}$ & $5.13723\times10^{-5}$ \\
4.7 & $1.00049$ & $1.00049$ & $2.67677\times10^{-5}$ \\
8 & $1.20836\times10^{-5}$ & $1.55344\times10^{-5}$ & $1.31082\times10^{-5}$ \\
12 & $1.00057$ & $1.00078$ & $8.60621\times10^{-6}$ \\
14 & $1.61438\times10^{-5}$ & $1.67080\times10^{-5}$ & $1.53611\times10^{-5}$ \\
\end{tabular}
\end{ruledtabular}
\end{table}

\begin{figure}[tp]
\includegraphics[width=\columnwidth]{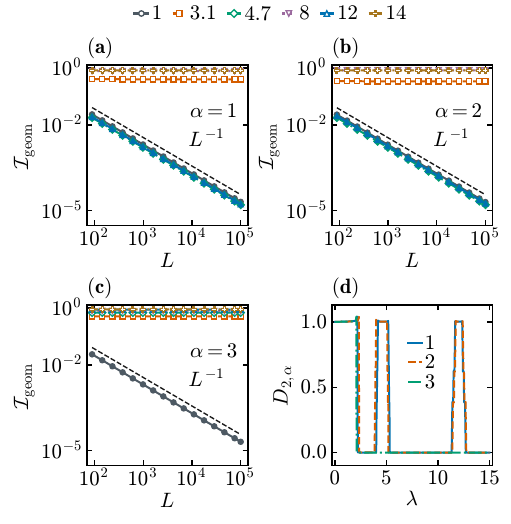}
\caption{Complete-band participation over four phases. (a--c) Geometric mean IPR at Eq.~\eqref{eq:points}, for \NumSizes\ lengths $L=89$--$\MaxCells$. Filled/open markers denote extended-type/localized-type scaling over these sizes. (d) Band-averaged $D_{2,\alpha}$ fitted to $L=10946,17711,28657,46368$ on the \ScalingScanPoints-point grid; shading spans the individual-phase fits. Dashed $L^{-1}$ guides in (a--c) are normalized to 1.8 times the $\lambda=1$, $L=100000$ IPR. Colors and symbols follow the parameter key above.}
\label{fig:scaling}
\end{figure}
At $\lambda=1,4.7,12$, both lower bands have an IPR that decreases approximately as $L^{-1}$ and an NPR that remains of order unity over the $\NumSizes$ sampled lengths up to $L=\MaxCells$. At $\lambda=3.1,8,14$, their IPR saturates and their NPR decreases with $L$ [Fig.~\ref{fig:scaling} and Table~\ref{tab:static}]. These samples establish two returns to extended-type scaling in each lower band. The third band has extended-type scaling at $\lambda=1$ and localized-type scaling at the other five samples.

The \ScalingScanPoints-point scan over $0\leq\lambda\leq15$ resolves the intervals between these samples [Fig.~\ref{fig:scaling}(d)]. Both lower-band exponents return close to one in the two high-participation windows, while the third-band exponent remains close to zero after its first crossover. For the representative samples, fits to $L=17711,28657,46368,75025$ and $L=46368,75025,100000$ assess the dependence on the fitting range.

Near the projected edge of the second window, the NPR quantiles at $\lambda=12.55$ have different size dependences, so the band-averaged exponents summarize a nonuniform response across eigenstates (Appendix~\ref{app:distribution}). This edge sample is used in the molecular-separation scans of Sec.~\ref{sec:interband}.

\subsection{Local propagation and finite-chain retention}
\label{sec:dynamics}
Wave-packet spreading probes the spatial consequences of localization~\cite{HiramotoAbe1988,Ketzmerick1997,Xu2020}, with related measurements in atomic and photonic lattices~\cite{Roati2008,Lahini2009,Luschen2018,Vaidya2023}. The three local states $|m_0,s\rangle$, $s=A,B,C$, were evolved independently at fixed Hamiltonian parameters, with $m_0=\lfloor L/2\rfloor$. Their equal incoherent average excites all three spectral bands.

Amplitudes $\psi_{m,s}(t)$ define the cell probabilities $p_m(t)=\sum_s|\psi_{m,s}(t)|^2$. The mean-square displacement and central retention are
\begin{equation}
\begin{aligned}
&M_2(t)=\sum_m(m-m_0)^2p_m(t),\\
&P(r,t)=\sum_{m=m_-}^{m_+}p_m(t).
\end{aligned}
\label{eq:dyn}
\end{equation}
The interval endpoints are $m_-=\max\{0,\lceil m_0-r/2\rceil\}$ and $m_+=\min\{L-1,\lfloor m_0+r/2\rfloor\}$. $\langle\cdots\rangle$ denotes the equal average over the three local launches and 32 phases. The root-mean-square (RMS) displacement $\sigma(t)=\sqrt{\langle M_2(t)\rangle}$ characterizes spreading from the launch cell in units of the cell spacing, with the average taken before the square root. The mean retention $\langle P(r,t)\rangle$ gives the fraction of probability remaining between $m_-$ and $m_+$. Both observables sum over all three orbitals and are invariant under intracell redistribution.

\begin{figure}[!htbp]
\includegraphics[width=\columnwidth]{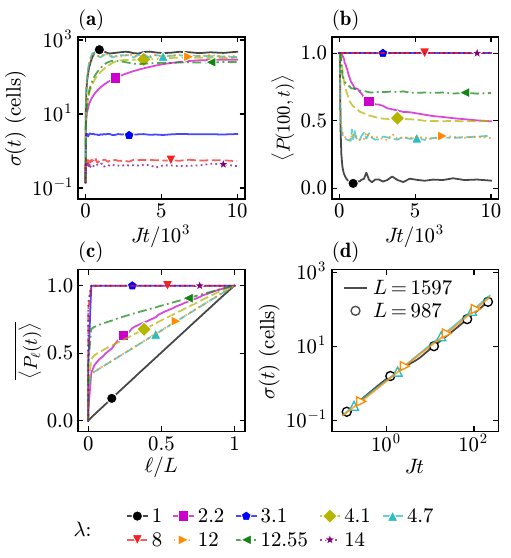}
\caption{Wave-packet dynamics from an equal incoherent mixture of the three local orbitals at nine $\lambda$ values and $L=1597$, averaged over 32 phases. (a) RMS displacement and (b) central-101-cell retention on linear time axes. (c) Long-time retention versus window fraction $\ell/L$, averaged over $5\times10^4\leq Jt\leq10^5$ on a separate stratified mesh; the central cell is at $\ell=1$. (d) Early spreading at $\lambda=1,4.7,12$ through $Jt=230$: lines $L=1597$, markers $L=987$. The parameter key below applies throughout.}
\label{fig:dyn}
\end{figure}

In Fig.~\ref{fig:dyn}(b), $P(100,t)$ includes cells $m_0-50$ through $m_0+50$, giving the probability retained in the central 101 cells. Panel (c) varies the number of included cells $\ell=r+1$ and uses $P_\ell(t)=P(\ell-1,t)$ for odd $1\leq\ell\leq L$, with $P_0(t)=0$.

The nine parameters in Fig.~\ref{fig:dyn} comprise the six values in Eq.~\eqref{eq:points} and $\lambda=2.2,4.1,12.55$. Evolution is computed to $Jt=\DynMaxTime$ at $L=\MainCells$ and $\CheckCells$, using centered modulation phases $x_m=2\pi\omega(m-m_0)+\theta$ to match the launch environments. The free-chain maximum speed $2J$ gives a boundary-arrival scale $Jt\sim L/4$. The boundary and two-length checks identify $Jt\leq230$ as the common early-time interval [Appendix~\ref{app:numerical}]; the long-time response is evaluated at fixed $L$.

Figures~\ref{fig:dyn}(a,b) contrast the reentrant samples $\lambda=4.7,12$ with the localized controls $\lambda=3.1,8,14$. At both reentrant points, $\sigma(t)$ grows to hundreds of cells while the central retention decreases substantially. The localized controls remain within a few cells and retain nearly all their probability in the central interval. The intermediate samples $\lambda=2.2,4.1,12.55$ also exhibit substantial spreading but retain a larger central component in the long-time regime.

Figure~\ref{fig:dyn}(d) shows approximately linear growth of $\sigma(t)$ at $\lambda=1,4.7,12$, with agreement between the two lengths throughout the early-time interval. The recovered propagation at the two reentrant points thus precedes finite-chain saturation. The saturation in panel (a) characterizes spreading on the finite chain; the long-time averaging protocol and sampling checks are given in Appendix~\ref{app:numerical}.

In Fig.~\ref{fig:dyn}(c), the localized controls retain nearly all their probability in a small central window. At both reentrant points, a sharp rise of retention at small $\ell$ is followed by continued growth as the window widens, indicating confined and spreading components. In the isolated-band limit, the equal orbital mixture assigns one third of its weight to each band; on a finite chain, extended states also contribute to the long-time central retention.

\begin{figure}[!htbp]
\includegraphics[width=\columnwidth]{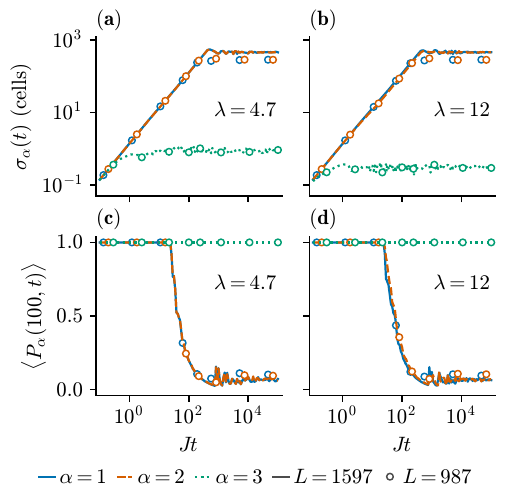}
\caption{Band-filtered dynamics from Eq.~\eqref{eq:bandlaunch}, averaged over 32 centered phases. Left and right columns use $\lambda=4.7$ and $12$, respectively. (a,b) RMS displacement and (c,d) probability retained in the central 101 cells, over $0<Jt\leq10^5$. Blue solid, orange dashed, and green dotted curves denote bands $\alpha=1,2,3$ at $L=1597$; open circles use $L=987$. Early spreading passes the boundary and two-length checks for $Jt\leq230$; long-time curves describe evolution on the finite chain.}
\label{fig:banddyn}
\end{figure}

Band conditioning resolves the spectral origin of the confined and spreading components in Fig.~\ref{fig:dyn}. With $S_0=\sum_s|m_0,s\rangle\langle m_0,s|$ and eigenenergies ordered from lowest to highest, the band projector and normalized initial density matrix are
\begin{equation}
\Pi_\alpha=\sum_{n=(\alpha-1)L+1}^{\alpha L}|E_n\rangle\langle E_n|,\qquad
\rho_{0,\alpha}=\frac{\Pi_\alpha S_0\Pi_\alpha}{\operatorname{Tr}(\Pi_\alpha S_0)}.
\label{eq:bandlaunch}
\end{equation}
The projector uses all $L$ full-model eigenstates in the selected band, including their interband admixture. The conditional state is generally mixed and evolves as $\rho_\alpha(t)=e^{-iHt}\rho_{0,\alpha}e^{iHt}$. Its phase-averaged initial RMS is below $10^{-3}$ cells for all six band--parameter combinations. The observables follow from Eq.~\eqref{eq:dyn} with $p_{m,\alpha}(t)=\sum_s\langle m,s|\rho_\alpha(t)|m,s\rangle$, taking the phase average before the RMS square root.

At both reentrant points, Fig.~\ref{fig:banddyn} separates the two spreading lower bands from the confined third band. By $Jt=230$, the lower-band RMS displacements reach hundreds of cells and their central retentions decrease, while the third band remains on the scale of one cell and retains nearly all its probability in the central interval.

The filtered ensembles satisfy the same boundary and two-length criteria over this early-time interval. In the long-time regime, the lower-band RMS saturation and retention depend on chain length, while the third-band width remains on the scale of one cell at both lengths for the specified phase ensemble. Phase-sampling sensitivity is reported in Appendix~\ref{app:banddyn}.

\section{Residual interband coupling and attenuation}
\label{sec:interband}
The full-ladder participation windows in Sec.~\ref{sec:bandparticipation} differ from the isolated-channel intervals of Sec.~\ref{sec:model}. The former use a finite-size NPR criterion, while the latter follow from the AA threshold. A matched-criterion comparison at fixed molecular levels and projected modulation isolates the changes induced by residual interband coupling.

\begin{figure}[!t]
\includegraphics[width=\columnwidth]{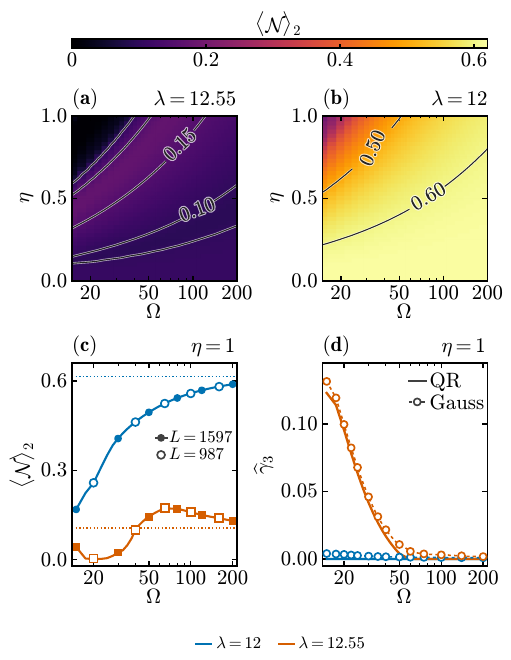}
\caption{Band-2 response to molecular separation and residual interband modulation. (a,b) Mean cell NPR near the projected edge ($\lambda=12.55$) and window center ($\lambda=12$), respectively, at $L=1597$, on $75\times41$ grids spanning $\Omega\in[15,200]$ and $\eta\in[0,1]$, with a common color scale. Contours: $0.10,0.15$ in (a), $0.50,0.60$ in (b). (c) Mean NPR at $\eta=1$: filled/open symbols, $L=1597/987$; dotted lines, decoupled references. Means in (a--c) cover the complete band and four phases. (d) Same band-middle-energy minimum-LE estimates at $\eta=1$: QR (solid), Gaussian determinants (dashed with open circles). Blue/orange in (c,d): $\lambda=12/12.55$. All $\Omega$ axes are logarithmic; (c,d) use 75 $\Omega$ samples.}
\label{fig:interband_response}
\end{figure}

\subsection{Interband self-energy}
\label{sec:finitegap}
Diagonal cancellation in the two candidate reentrant windows of Fig.~\ref{fig:model}(c) leaves finite off-diagonal modulation. Orthogonality of $\widetilde Q=R^TQR$ gives
\begin{equation}
q_\alpha^2+\sum_{\beta\ne\alpha}|\widetilde Q_{\alpha\beta}|^2=1.
\label{eq:modulationbudget}
\end{equation}
At $q_\alpha=0$, the off-diagonal coefficients satisfy $\sum_{\beta\ne\alpha}|\widetilde Q_{\alpha\beta}|^2=1$. Their effect on a given band depends on the modulation amplitude relative to the molecular level separations.

Let $\Delta_\alpha=\min_{\beta\ne\alpha}|\mu_\alpha-\mu_\beta|$ denote the molecular level separation. The operator norm of the longitudinal hopping and quasiperiodic terms is at most $2+\lambda$. The energy-dependent self-energy $\Sigma_\alpha(E)$ is the operator correction generated by eliminating the other molecular channels. For $|E-\mu_\alpha|\leq2+\lambda$ and $\Delta_\alpha>4+2\lambda$, the standard Feshbach projection~\cite{Feshbach1958} yields, for this ladder, the interband self-energy bound
\begin{equation}
\|\Sigma_\alpha(E)\|\leq
\frac{\lambda^2}{\Delta_\alpha-4-2\lambda}.
\label{eq:bound}
\end{equation}
At fixed $\lambda$, increasing $\Delta_\alpha$ lowers this upper bound on the residual perturbation, bringing the effective channel operator closer to the isolated AA form. For finite molecular separation, $\Sigma_\alpha(E)$ modifies both the spatial modulation and longitudinal propagation within the channel. Its spatial structure determines how the full-model attenuation departs from the projected LE in Eq.~\eqref{eq:LEapprox}.

For molecular separations large compared with the hopping and modulation scales, a second-order effective-Hamiltonian expansion~\cite{Bravyi2011} gives the leading local correction to the onsite potential
\begin{equation}
\delta V_{\alpha,m}\simeq\lambda^2\cos^2x_m
\sum_{\beta\ne\alpha}\frac{|\widetilde Q_{\alpha\beta}|^2}{\mu_\alpha-\mu_\beta}.
\label{eq:second}
\end{equation}
The identity $\cos^2x_m=[1+\cos(2x_m)]/2$ separates a uniform energy shift from a second harmonic. The latter is not proportional to $q_\alpha$ and survives at the two lower bands' orbital cancellation points. Its magnitude $A_{2,\alpha}$ depends on $\lambda^2$, the interband matrix elements, and the molecular level separations.

At $\Omega=100$, $A_{2,1}(4.7)=0.0877944$ and $A_{2,1}(12)=0.238345$. The ratios $A_{2,\alpha}(12)/A_{2,\alpha}(4.7)$ are approximately $2.71$ and $2.74$ for $\alpha=1,2$, respectively.

A weak shift of the critical projected amplitude on the scale $A_{2,\alpha}$ would correspond to a local parameter scale $A_{2,\alpha}/|s_\alpha|$, where $s_\alpha=d(\lambda q_\alpha)/d\lambda$. At $\lambda=4.7,12$, $|s_1|\simeq3.91,3.99$ and $|s_2|\simeq4.08,4.03$, respectively, giving scale ratios $2.67$ and $2.77$ for the two lower channels.

Elimination of an antisymmetrically coupled channel generates a squared potential exactly in flat-band models~\cite{Bodyfelt2014,Danieli2015}. For the three dispersive channels, Eq.~\eqref{eq:second} gives the leading local correction; energy dependence and nonlocal terms also enter the full self-energy. These terms can modify the LE and window edges relative to the isolated-channel criterion $|\lambda q_\alpha|=2$.

The diagnostic interpolation
\begin{equation}
\widetilde Q_\eta=\operatorname{diag}(q_\alpha)
+\eta[\widetilde Q-\operatorname{diag}(q_\alpha)],\quad 0\leq\eta\leq1,
\label{eq:eta}
\end{equation}
connects the projected channels at $\eta=0$ to the full model at $\eta=1$, varying the interband modulation at fixed molecular levels and projected onsite potentials. In the large-separation regime, the leading self-energy in Eq.~\eqref{eq:second} scales as $\eta^2$. Increasing the molecular scale $\Omega$ enlarges the molecular level separations. These complementary controls reveal different sensitivities at the second-window center and near the projected edge of band 2.

\begin{table*}[t]
\caption{Matched-criterion endpoint comparison at $\Omega=100$. The listed interiors contain consecutive sampled controls at which at least $90\%$ of states in each lower band, pooled over the same four phases, satisfy $\Ncell>0.1$ at both $L=987,1597$. The shifts $\delta\lambda_{\rm l,r}$ are the left/right endpoints at $\eta=1$ minus those at $\eta=0$, and the width change is $\Delta W=\delta\lambda_{\rm r}-\delta\lambda_{\rm l}$. Adjacent failing controls are $0.005$ below and above each interior. Their brackets give the displayed grid-resolution ranges; the sampling protocol is given in Appendix~\ref{app:matchedendpoints}.}
\label{tab:matchedendpoints}
\setlength{\tabcolsep}{4pt}
\begin{tabular}{cccccc}
\toprule
Window & $\eta=0$ interior & $\eta=1$ interior & $10^3\delta\lambda_{\rm l}$ & $10^3\delta\lambda_{\rm r}$ & $10^3\Delta W$\\
\midrule
1 & $[4.13000,5.11500]$ & $[4.18000,5.07000]$ & $[45.0000,55.0000]$ & $[-50.0000,-40.0000]$ & $[-105.000,-85.0000]$ \\
2 & $[11.5850,12.5000]$ & $[11.7200,12.3650]$ & $[130.000,140.000]$ & $[-140.000,-130.000]$ & $[-280.000,-260.000]$ \\
\bottomrule
\end{tabular}
\end{table*}

\subsection{Energy-resolved attenuation}
\label{sec:transfer}
For the full trimer ladder ($\eta=1$), the block Schr\"odinger equation has transfer matrix
\begin{equation}
\begin{gathered}
\begin{pmatrix}\boldsymbol\psi_{m+1}\\\boldsymbol\psi_m\end{pmatrix}
=\mathsf T_m(E)\begin{pmatrix}\boldsymbol\psi_m\\\boldsymbol\psi_{m-1}\end{pmatrix},\\
\mathsf T_m(E)=\begin{pmatrix}E I-\Omega K-\lambda\cos x_m Q&-I\\I&0\end{pmatrix},
\end{gathered}
\label{eq:transfer}
\end{equation}
where $\boldsymbol\psi_m$ contains the three orbital amplitudes and $I$ is the $3\times3$ identity. Orthogonal--triangular (QR) factorization gives finite-length LE estimates using standard orthogonalization methods~\cite{Benettin1980,MacKinnonKramer1983}. The six infinite-length exponents form pairs $\pm\gamma_j$, with $\gamma_1\geq\gamma_2\geq\gamma_3\geq0$. The index $j$ orders exponents at the chosen energy and differs from the molecular band label $\alpha$. The smallest nonnegative LE, $\gamma_3$, measures the weakest longitudinal attenuation: $\gamma_3>0$ gives exponential attenuation in every channel, while $\gamma_3=0$ permits a channel without exponential attenuation even when the other two nonnegative exponents are positive. Extended and critical states can both have vanishing minimum LE. Participation scaling in Sec.~\ref{sec:bandparticipation} and propagation in Sec.~\ref{sec:dynamics} characterize their spatial behavior over the sampled sizes and times.

Repeated QR factorization accumulates the logarithmic stretch along each orthogonal direction:
\begin{equation}
\mathsf T_m\mathsf U_{m-1}=\mathsf U_m\mathsf R_m,\qquad
\widehat\gamma_j^{\rm QR}=\frac1{N_{\rm QR}}\sum_m\log|\mathsf R_{m,jj}|.
\label{eq:QR}
\end{equation}
In Eq.~\eqref{eq:QR}, $\mathsf U_m$ is orthogonal and $\mathsf R_m$ is upper triangular; $N_{\rm QR}$ counts retained cells after a discarded transient. The estimates average over two randomly chosen starting phases kept fixed throughout the scan. For each $(\lambda,\Omega)$ pair, the energy is selected separately as the middle-ranked eigenenergy of band 2 in an open chain. This diagnostic resolves attenuation at the selected energy, while the NPR averages cover the complete band.

The attenuation estimates are cross-checked using logarithmic determinants at the same energies through the determinant--transfer relation~\cite{Molinari2003SpectralDuality,Deninger2011}. Closed-walk moments yield Gaussian-quadrature evaluations of unbiased and biased logarithmic determinants, whose combination gives the minimum-LE estimates in Fig.~\ref{fig:interband_response}(d). At matched moment degree, Gaussian quadrature improves convergence relative to the truncated logarithmic series in the sampled benchmarks [Fig.~\ref{fig:gauss_method}]. Comparing them with QR, together with transfer-length and quadrature-order refinements, tests the numerical stability of the attenuation trends. The determinant construction, corresponding truncated logarithmic series, and convergence checks are given in Appendices~\ref{app:determinant}--\ref{app:walkproof} and~\ref{app:numerical}.

\subsection{Sensitivity to residual interband modulation}
\label{sec:couplingsensitivity}
At fixed $u$, $v$, and $\Omega$, the interpolation in Eq.~\eqref{eq:eta} restores interband modulation without changing the molecular levels or projected onsite potentials. At $\Omega=100$, the same two-band NPR criterion, phases, and lengths as in Sec.~\ref{sec:bandparticipation} give the sampled interiors in Table~\ref{tab:matchedendpoints}. Both models use a common control mesh with $0.005$ spacing at their endpoint brackets (Appendix~\ref{app:matchedendpoints}).

Under this finite-sample criterion, restoring the full off-diagonal modulation shifts both left endpoints upward and both right endpoints downward. The propagated width-change brackets are negative for both windows and resolve a larger absolute contraction of the second.

The sampled width-reduction ratio is approximately $2.84$, with the grid brackets in Table~\ref{tab:matchedendpoints} giving $[2.47,3.30]$. Both $A_{2,\alpha}/|s_\alpha|$ ratios in Sec.~\ref{sec:finitegap} fall within this range, showing semiquantitative consistency with the stronger contraction of the second window. This comparison uses representative controls; endpoint shifts depend on the criterion's response to the second harmonic, the slopes at the edges, and the full self-energy.

The edge sample $\lambda=12.55$, introduced in Sec.~\ref{sec:bandparticipation}, probes this influence close to the isolated-channel threshold in band 2. The center sample $\lambda=12$ provides a comparison with a more strongly suppressed projected onsite potential.

The $\eta$ scans in Fig.~\ref{fig:interband_response}(a,b) distinguish the edge response from that at the center. At the edge, restoring the interband modulation can either suppress or enhance participation relative to the decoupled limit, depending on the molecular separation. The center has a higher mean NPR throughout the sampled grid and retains a broad region close to its decoupled reference [panel (b)].

At fixed $u$ and $v$, increasing $\Omega$ leaves $q_\alpha$ unchanged and enlarges the molecular level separations. Within the separation regime of Eq.~\eqref{eq:bound}, this lowers the upper bound on the residual self-energy. In the large-separation regime, the leading interband self-energy scales as $\eta^2/\Omega$ [Eq.~\eqref{eq:second}]. This scan also rescales the intracell hoppings; each chosen $\Omega$ defines a separate joint $\lambda$ path with fixed hoppings. In the full model, the center NPR increases toward its decoupled reference as $\Omega$ grows, whereas the edge mean remains nonmonotone and can exceed its decoupled value [Fig.~\ref{fig:interband_response}(c)].

At the edge, an increase in the band mean coexists with a broad distribution of eigenstate participation. At $\Omega=50$ and $L=1597$, the phase-pooled mean NPR is $0.144712$, and $46.2899\%$ of states satisfy $\Ncell>0.1$. The median decreases with chain length at this sample. Eigenstate statistics and sampling checks are given in Appendix~\ref{app:walknumerics}.

At the selected band-middle energies, the QR minimum remains close to zero at the center and decreases toward zero at the edge as the molecular separation grows [Fig.~\ref{fig:interband_response}(d)]. The decreasing edge attenuation at a selected energy coexists with the nonmonotone mean NPR over the complete band in panel (c). Gaussian determinant extraction reproduces the edge attenuation trend with a finite-node offset from QR; weak-attenuation samples retain quadrature-order sensitivity. The series comparison and convergence checks are given in Appendix~\ref{app:walknumerics}.

\section{Control-path dependence}
\label{sec:scope}
The residual-coupling scans in Sec.~\ref{sec:interband} retain the reference relation $u=v$. Independent variation of the detuning $u$ and modulation amplitude $v$ probes the persistence of the two high-participation windows along nearby monotone paths, with all hoppings fixed. The molecular scale remains $\Omega=100$ throughout this comparison.

For each complete molecular band, the phase-pooled fraction above the NPR threshold is
\begin{align}
f_\alpha(u,v;L)&=\frac1{M_\theta L}\sum_{\ell=1}^{M_\theta}\sum_{n\in\alpha}
\mathbf1[\Ncell_{n,\ell}>0.1],\nonumber\\
f_{\min}(u,v;L)&=\min\{f_1,f_2\}.
\label{eq:fraction}
\end{align}
The index $\ell$ labels the $M_\theta=4$ sampled phases, and $\mathbf1$ is the indicator function. The pooled criterion $f_{\min}\geq0.9$ is the same two-band threshold used in Sec.~\ref{sec:bandparticipation}.

\begin{figure}[tp]
\includegraphics[width=\columnwidth]{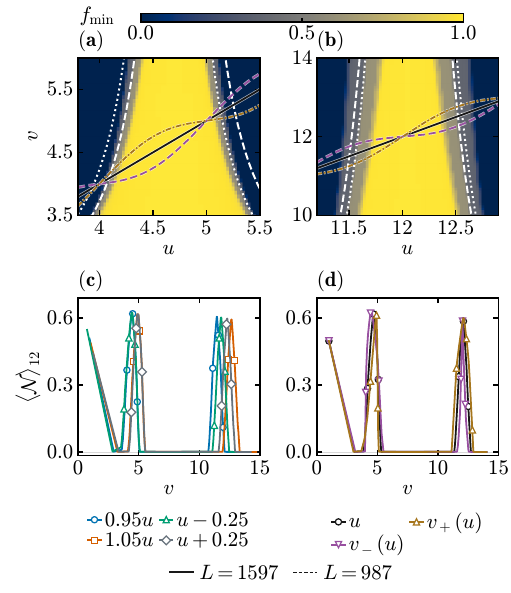}
\caption{Control plane and path dependence at $\Omega=100$. (a,b) Phase-pooled lower-band fraction $f_{\min}$ [Eq.~\eqref{eq:fraction}] in the first/second windows at $L=1597$. Black solid, purple dashed, and brown dash-dotted paths denote $v=u,v_-,v_+$. White dashed/dotted curves give $|vq_\alpha(u)|=2$ for bands 1/2. (c,d) Mean cell NPR pooled over both lower bands and four phases versus the modulation amplitude $v$: (c) four affine perturbations; (d) the reference and two nonlinear paths. Colors and open symbols distinguish paths; the legends give $v(u)$. Solid/thin dashed lines denote $L=1597/987$. Independent samples cover $1\leq u\leq14$, with finer sampling near both windows.}
\label{fig:paths}
\end{figure}

Figures~\ref{fig:paths}(a,b) show two high-participation regions extending to both sides of the reference line $v=u$. Their positions in $u$ follow the two orbital-cancellation regions identified in Sec.~\ref{sec:mechanism}. The white curves $|vq_\alpha(u)|=2$ mark the isolated-channel thresholds; high full-model fractions occur where both lower-band projected modulation amplitudes are suppressed.

The five straight paths, $v=u$, $v=0.95u$, $v=1.05u$, and $v=u\pm0.25$, test changes in the relative growth rate and fixed amplitude offsets. Two nonlinear paths are
\begin{equation}
v_\pm(u)=u\pm\frac14\sin(\pi u),\qquad 0\leq u\leq15.
\label{eq:curvedpath}
\end{equation}
Their deviations from the reference path satisfy $|v_\pm-u|\leq0.25$, and $v_\pm'(u)\geq1-\pi/4>0$. Thus both detuning and modulation amplitude increase monotonically despite the path curvature.

For the path profiles, the mean cell NPR pooled over the two lower bands is $\langle\Ncell\rangle_{12}=(\langle\Ncell\rangle_1+\langle\Ncell\rangle_2)/2$, with each band mean averaging over all eigenstates and four phases. Figure~\ref{fig:paths}(c,d) plots this mean against the actual modulation amplitude $v$ over the sampled range $1\leq u\leq14$. Panels (c) and (d) compare the four affine perturbations and the reference with two nonlinear paths, respectively, on common NPR and $v$ scales. Both $u$ and $v$ vary during these scans; monotonicity of $v(u)$ preserves the order of the sampled parameter points.

Along all seven paths, the weak-modulation decrease of the mean NPR is followed by two recoveries at both $L=987$ and $1597$, separated by a low-participation interval. Changes in slope, offset, and curvature shift the profiles along $v$, while both high-participation windows persist. Table~\ref{tab:paths} lists the consecutive sampled $u$ intervals satisfying $f_{\min}\geq0.9$ in both lower bands at both lengths for the four-phase ensemble; sampling and threshold refinement are specified in Appendix~\ref{app:staticdetails}.

\section{Discussion and conclusions}
\label{sec:conclusion}
The joint detuning--amplitude path creates a competition between increasing microscopic modulation and decreasing molecular-channel projections near avoided crossings. The two lower molecular channels approach diagonal cancellation while all hopping amplitudes and modulation signs remain fixed. The orbital-weight sum rule links their paired cancellations to strong modulation in the third band. Complete-band participation scaling and early propagation support recovery in both lower bands over the studied sizes and times. At the reentrant samples, local dynamics combine a confined third-band component with spreading lower-band components, which also contribute to central retention in finite chains.

Diagonal cancellation leaves finite interband scattering. Eliminating the other molecular channels produces a self-energy whose magnitude depends on the molecular level separations. In the large-separation regime, its leading local contribution contains a second harmonic that survives at the cancellation points. At the reference molecular scale, the coupled--decoupled comparison holds the molecular levels and projected potentials fixed. Under the same two-band finite-size participation criterion, restoring the full off-diagonal modulation shifts both sampled endpoints inward in each window, with a larger absolute contraction of the second.

The two high-participation regions remain accessible along all seven tested nearby monotone paths, despite changes in the relative growth of detuning and modulation. Molecular-separation scans distinguish the second-window center, whose participation approaches the decoupled reference, from a near-edge sample with nonmonotone band-averaged participation. At this near-edge sample, decreasing attenuation at selected energies accompanies the nonmonotone complete-band participation, reflecting the different spectral content of the two observables. The reported endpoints delimit sampled windows for the specified phase ensemble, chain lengths, and participation criterion.

\begin{acknowledgments}
This work was supported by the National Natural Science Foundation of China under Grants No. 11974205, No. 61727801, and No. 62131002, and by the Key Research and Development Program of Guangdong Province (2018B030325002).
\end{acknowledgments}

\section*{Data availability}
Detailed numerical records for the numerical-protocol appendix are provided as comma-separated-value (CSV) files in the ancillary material~\cite{SupplementalMaterial}. The remaining supporting data and code are available from the authors upon reasonable request.
\appendix
\section{Supplementary determinant evaluation}
\label{app:determinant}
\label{sec:walk}
This appendix presents the biased-determinant framework and its series and Gaussian evaluations. Appendix~\ref{app:walkconstruction} supplies the explicit closed-walk construction and scalar benchmarks, while Appendix~\ref{app:walkproof} gives the determinant factorization proof and conditions for rigorous error control. Appendix~\ref{app:distribution} analyzes participation distributions and fitting-range dependence. For numerical protocols and convergence checks, see Appendix~\ref{app:numerical}.

A biased logarithmic determinant provides a supplementary estimate of the minimum LE defined in Sec.~\ref{sec:transfer}. Closed-walk moments retain the irrational modulation and ordered interband matrix products. The logarithmic series and Gaussian quadrature approximate the same spectral integral at matched moment degree, providing same-energy comparisons with transfer estimates. Node and phase sensitivities are assessed separately from QR length dependence.

\subsection{Isolated-channel complex-phase benchmark}
\label{sec:complexbenchmark}
For $a_\alpha=vq_\alpha(u)$ and $e_\alpha=E-\mu_\alpha(u)$, continuation of the phase to $x+i\epsilon$ gives, on the real spectrum of the projected AA operator,
\begin{equation}
\gamma_\alpha^{(0)}(e_\alpha,\epsilon)
=\max\{0,\mathcal B_\alpha^{(0)}+|\epsilon|\},\qquad
\mathcal B_\alpha^{(0)}=\log\frac{|a_\alpha|}{2}.
\label{eq:avila_main}
\end{equation}
The imaginary phase $\epsilon$ is measured in radians, so the nonzero slope is one. A negative intercept gives a zero-exponent strip, the subcritical regime; the strip closes at the critical point. A positive intercept gives positive real-axis growth, the supercritical regime~\cite{Avila2015}. Exact cancellation gives the free-chain limit $\mathcal B_\alpha^{(0)}=-\infty$.

At an arbitrary real energy, the scalar expression is $\max\{\gamma_\alpha^{(0)}(e_\alpha,0),\mathcal B_\alpha^{(0)}+|\epsilon|\}$. Its first entry vanishes on the subcritical projected spectrum. For the coupled ladder, the higher-dimensional cocycle~\cite{AvilaJitomirskayaSadel2014} is treated through the determinant identity below.

\subsection{Biased determinant extraction}
An auxiliary hopping bias, introduced as an imaginary vector potential by Hatano and Nelson~\cite{HatanoNelson1996}, provides a determinant diagnostic for the Hermitian localization problem. For $\vartheta=2\pi\omega$, define $(U\psi)_m=\psi_{m+1}$ and $(Z\psi)_m=e^{ix_m}\psi_m$, with $UZ=e^{i\vartheta}ZU$. Continuing the phase as $x_m\mapsto x_m+i\epsilon$ and introducing a hopping bias $h\geq0$ gives
\begin{align}
F_{\epsilon,h}={}&EI-\Omega K-e^hUI-e^{-h}U^{-1}I\nonumber\\
&-\frac{\lambda}{2}(e^{-\epsilon}Z+e^\epsilon Z^{-1})Q.
\label{eq:Fwalk}
\end{align}
The physical Hamiltonian has $h=0$ and $\epsilon=0$. For $A=\sum_{s,k}A_{s,k}Z^kU^s$, the canonical trace per cell is
\begin{equation}
\tau(A)=\operatorname{tr}A_{0,0},\qquad\tau(I)=3.
\label{eq:celltrace}
\end{equation}
This is the phase-averaged onsite orbital trace. Zero cell displacement selects the onsite block, while phase averaging selects its zero Fourier component.
The logarithmic Fuglede--Kadison determinant~\cite{FugledeKadison1952,deLaHarpe2013} is
\begin{equation}
\mathcal D_{\epsilon,h}=\log\Delta(F_{\epsilon,h})
=\tfrac12\tau\log(F_{\epsilon,h}^{\dagger}F_{\epsilon,h}),
\label{eq:Dwalk}
\end{equation}
with decreasing positive regularization at zero singular values. Specializing Deninger's cocycle determinant theorem (Theorem~9 and Eq.~(32) of Ref.~\onlinecite{Deninger2011}) to the transfer operator gives
\begin{equation}
\mathcal D_{\epsilon,h}=\sum_{j=1}^3\max\{\gamma_j(E,\epsilon),h\}.
\label{eq:Dchannels}
\end{equation}
Here $\gamma_j(E,\epsilon)$ are the three nonnegative infinite-length exponents of the complex symplectic transfer matrix. Under the separation condition
\begin{equation}
\gamma_3(E,\epsilon)<h<\gamma_2(E,\epsilon),
\label{eq:separator}
\end{equation}
subtracting the biased identity from its unbiased counterpart gives
\begin{equation}
\gamma_3(E,\epsilon)=\mathcal D_{\epsilon,0}-\mathcal D_{\epsilon,h}+h.
\label{eq:extract}
\end{equation}
Molinari's finite-matrix spectral-duality construction identifies transfer exponents through the bias dependence~\cite{Molinari2003SpectralDuality}, including the smallest exponent in Sec.~III.D of Ref.~\onlinecite{Molinari2009NonHermitianSpectra}. Equations~\eqref{eq:Dchannels} and~\eqref{eq:extract} express this extraction in the infinite ergodic-operator setting. The determinants are evaluated from local closed-walk moments at the irrational frequency.

\subsection{The logarithmic series and its Gaussian evaluation}
\begin{figure}[t]
\includegraphics[width=\columnwidth]{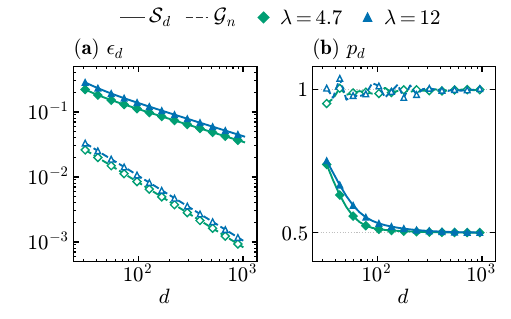}
\caption{Full-operator numerical benchmark at a matched moment degree. (a) Discrepancy $\epsilon_d=|\widehat D_d-D^{\rm QR}|$, where $\widehat D_d$ is the series or Gaussian estimate and $D^{\rm QR}$ is the sum of QR exponents, at first-band energies for $\lambda=4.7,12$: series $\mathcal S_d$ (solid) and Gaussian rule $\mathcal G_n$ (dashed), with $d=2n-1$. (b) Adjacent-budget orders $p_d=-\Delta\log\epsilon_d/\Delta\log d$, shown at geometric midpoints; dotted guides mark $1/2$ and $1$. These orders refer to the sampled budgets and finite-length QR reference.}
\label{fig:gauss_method}
\end{figure}

To evaluate the determinant, introduce a positive diagonal preconditioner $\mathsf W$ in the molecular basis and a norm bound $\mathcal R\geq\|\widehat F\|$, with
\begin{align}
\widehat F&=\mathsf W^{1/2}R^TF_{\epsilon,h}R\mathsf W^{1/2},\nonumber\\
X&=\widehat F^\dagger\widehat F/\mathcal R^2,\quad
\mathscr K=I-X,\quad m_k=\tau(\mathscr K^k).
\label{eq:precondition}
\end{align}
The logarithmic series is
\begin{equation}
\mathcal D_{\epsilon,h}=3\log\mathcal R-\log\det\mathsf W
-\frac12\sum_{k=1}^{\infty}\frac{m_k}{k}.
\label{eq:mainseries}
\end{equation}
The positive-operator logarithmic series applies at arbitrary modulation amplitude. The canonical trace selects paths with zero total cell displacement and Fourier index. Their ordered matrix products retain interband coupling and phases from the irrational translation. The scalar relation between magnetic-translation traces and closed lattice walks is established in the Hofstadter literature~\cite{Ouvry2016}. The present weights include ordered orbital matrices, with coefficients and low-order checks given in Appendix~\ref{app:walkconstruction}.

Orbital-resolved closed-walk moments also determine a spectral integral. For each molecular orbital $a$, let $\nu_a$ be the unit-mass local spectral measure of the positive operator $X$ at the cell origin, including phase averaging. Since $0\leq X\leq I$,
\begin{align}
m_k&=\sum_{a=1}^3\int_0^1(1-x)^k\,d\nu_a(x),\nonumber\\
\tau\log X&=\sum_{a=1}^3\int_0^1\log x\,d\nu_a(x),
\label{eq:momentmeasure}
\end{align}
with the same decreasing regularization at $x=0$. The scalar $m_k$ sums the three orbital-resolved moments. The series in Eq.~\eqref{eq:mainseries} integrates $\log x$ through its expansion in $(1-x)^k$. Alternatively, an $n$-node Gaussian rule matches each local measure's moments through degree $2n-1$ and evaluates the logarithm at its nodes:
\begin{align}
\mathcal G_n={}&3\log\mathcal R-\log\det\mathsf W\nonumber\\
&+\frac12\sum_{a=1}^{3}\sum_{\ell=1}^{n}w_{a,\ell}\log x_{a,\ell}.
\label{eq:maingauss}
\end{align}
Lanczos Jacobi matrices determine the positive quadrature weights and nodes~\cite{GolubWelsch1969,GolubMeurant2010,Zimmerling2025}. Quadrature of matrix-function bilinear forms, including logarithms for determinant evaluation, is an established numerical method~\cite{BaiFaheyGolub1996,Ubaru2017}. Uniformly sampled phases enter a single direct-sum Lanczos inner product for each orbital. Moment matching is exact for polynomials through degree $2n-1$, while the logarithmic integral is approximated. The comparison of $\mathcal G_n$ and $\mathcal S_{2n-1}$ holds the measure, preconditioner, and phase mesh fixed and measures accuracy at equal moment degree.

\subsection{Independent transfer benchmark and convergence}
Figure~\ref{fig:gauss_method} compares 26 matched degrees $d=31$--$1023$ at 512 phases, with $D^{\rm QR}=\sum_{j=1}^3\widehat\gamma_j^{\rm QR}$ at $\epsilon=h=0$. At 512 nodes, the Gaussian determinant discrepancies at $\lambda=4.7,12$ are smaller than the matched-degree series discrepancies by factors $42.2977$ and $41.0137$, respectively. Energy samples, bias margins, and node, phase, and transfer refinements are given in Appendix~\ref{app:walknumerics}. Weak-attenuation samples retain node sensitivity. Certified exponent bounds require two-sided determinant control and the separation condition in Appendix~\ref{app:walkproof}.

\section{Closed-walk coefficients and scalar benchmarks}
\label{app:walkconstruction}
\subsection{Explicit coefficients and convergent series}
An invertible positive diagonal preconditioner in the molecular basis is
\begin{equation}
\begin{gathered}
\mathsf W=\operatorname{diag}(p_\alpha),\quad
p_\alpha=\frac{1}{\max\{1,|E-\mu_\alpha|\}},\\
\widehat F=\mathsf W^{1/2}R^TF_{\epsilon,h}R\mathsf W^{1/2}.
\end{gathered}
\end{equation}
The five nonzero coefficients of $\widehat F=\sum C_{s,k}Z^kU^s$ are
\begin{align}
C_{0,0}&=\mathsf W(EI-\operatorname{diag}\mu_\alpha),\nonumber\\
C_{\pm1,0}&=-e^{\pm h}\mathsf W,\nonumber\\
C_{0,\pm1}&=-\frac{\lambda}{2}e^{\mp\epsilon}
\mathsf W^{1/2}\widetilde Q\mathsf W^{1/2}.
\label{eq:fivecoeff}
\end{align}
Taking $\mathsf W=I$ gives the unscaled construction. For either choice, an explicit admissible norm bound is
\begin{align}
\mathcal R={}&\|C_{0,0}\|+2\cosh h\,\|\mathsf W\|\nonumber\\
&+\lambda\cosh\epsilon\,\|\mathsf W^{1/2}\widetilde Q\mathsf W^{1/2}\|.
\label{eq:Rwalk}
\end{align}
This $\mathcal R$ is a computational scale, distinct from the molecular rotation $R$. The definitions of $X$, $\mathscr K$, and $m_n$ in Eq.~\eqref{eq:precondition} give $0\leq X\leq I$, $0\leq\mathscr K\leq I$, and $m_n\geq0$. The compensating term $-\log\det\mathsf W$ in Eq.~\eqref{eq:mainseries} makes the positive-contraction expansion valid at arbitrary modulation amplitude.

Every coefficient is a finite expression in the original irrational frequency and the microscopic parameters. In the convention of Eq.~\eqref{eq:celltrace}, multiplication and adjunction obey
\begin{align}
(AB)_{s,k}&=\sum_{u,v}A_{u,v}B_{s-u,k-v}
e^{i\vartheta u(k-v)},\label{eq:twist}\\
(A^{\dagger})_{-s,-k}&=e^{i\vartheta sk}A_{s,k}^{\dagger}.
\end{align}
If $\mathscr K=\sum K_{s,k}Z^kU^s$, the explicit closed-walk coefficient is
\begin{align}
m_n
=\!\!\sum_{\substack{\sum_j s_j=0\\\sum_j k_j=0}}
e^{i\vartheta\sum_{i<j}s_i k_j}
\operatorname{tr}(K_{s_1,k_1}\cdots K_{s_n,k_n}).
\label{eq:closedpaths}
\end{align}
The zero-displacement and zero-Fourier-index constraints select closed paths, retaining their ordered phases and signed matrix amplitudes. The support of $\mathscr K$ lies in $|s|+|k|\leq2$. Recursive coefficient multiplication at the irrational frequency gives the low-order identities
\begin{align}
m_1&=3-\mathcal R^{-2}\sum_{s,k}\|C_{s,k}\|_{\mathrm F}^2,\nonumber\\
m_2&=\sum_{s,k}\|K_{s,k}\|_{\mathrm F}^2.
\label{eq:lowmoments}
\end{align}
Here $\|\cdot\|_{\mathrm F}$ denotes the Frobenius norm.

Denote the truncation of Eq.~\eqref{eq:mainseries} at order $N$ by $\mathcal S_N$. Positivity gives
\begin{equation}
\mathcal D_{\epsilon,h}\leq\mathcal S_{N+1}\leq\mathcal S_N,
\qquad \mathcal S_N\downarrow\mathcal D_{\epsilon,h}.
\label{eq:monotone}
\end{equation}
If a bound $\|\mathscr K\|\leq\rho<1$ is available, the tail satisfies
\begin{equation}
0\leq\mathcal S_N-\mathcal D_{\epsilon,h}
\leq\frac{3\rho^{N+1}}{2(N+1)(1-\rho)}.
\label{eq:tailwalk}
\end{equation}
The strict-contraction domain and conditions for bounding the determinant difference are given in Appendix~\ref{app:walkproof}.

\subsection{Gaussian quadrature from the same moments}
The local spectral measures and Gaussian estimator are defined in Eqs.~\eqref{eq:momentmeasure} and~\eqref{eq:maingauss}. The $n$-node rule matches moments through degree $2n-1$, with positive nodes and weights. In exact arithmetic with the canonical moments,

\begin{equation}
\mathcal D_{\epsilon,h}\leq\mathcal G_{n+1}\leq\mathcal G_n,
\qquad \mathcal G_n\downarrow\mathcal D_{\epsilon,h}.
\label{eq:Gaussbound}
\end{equation}
The upper-bound direction follows from the negative even derivatives of $\log x$. Its resolvent representation and nested Lanczos spaces give monotonicity~\cite{GolubMeurant2010,Zimmerling2025}. Lanczos recursion forms the Jacobi matrices. The estimates $\mathcal G_n$ and $\mathcal S_{2n-1}$ use the same maximum moment degree.

\subsection{The signed single-band quantity \texorpdfstring{$\mathcal B_\alpha^{(0)}$}{B}}
For $a_\alpha=\lambda q_\alpha$ and $e_\alpha=E-\mu_\alpha$, use the signed intercept and projected-spectrum benchmark in Eq.~\eqref{eq:avila_main}. For $a_\alpha\ne0$, choose $\epsilon_\alpha>\max\{0,-\mathcal B_\alpha^{(0)}\}$. With the scalar operator
\begin{equation}
f=e_\alpha-U-U^{-1}
-\frac{a_\alpha}{2}(e^{-\epsilon_\alpha}Z+e^{\epsilon_\alpha}Z^{-1}),
\end{equation}
the same construction now has $\tau(I)=1$ and gives
\begin{equation}
\mathcal B_\alpha^{(0)}
=\log r-\epsilon_\alpha
-\frac12\sum_{n=1}^{\infty}\frac{\tau[(I-f^{\dagger}f/r^2)^n]}{n},
\label{eq:Bseries}
\end{equation}
where $r\geq|e_\alpha|+2+|a_\alpha|\cosh\epsilon_\alpha$. Replacing the scalar logarithmic series by its Gaussian rule gives an alternative evaluation of $\mathcal B_\alpha^{(0)}$. A negative finite upper bound proves its negative sign under the stated spectral and arithmetic conditions. At exact cancellation $a_\alpha=0$, the projected chain is free and $\mathcal B_\alpha^{(0)}=-\infty$ is defined by the limiting expression.

With the band index suppressed, $k=I-f^\dagger f/r^2$, $M_1=\tau(f^\dagger f)$, and $M_2=\tau[(f^\dagger f)^2]$, the first two coefficients are
\begin{align}
M_1={}&e^2+2+\frac{a^2}{2}\cosh(2\epsilon),\nonumber\\
M_2={}&M_1^2+8e^2+2e^2a^2\cosh^2\epsilon+2\nonumber\\
&+\frac{a^4}{8}+2a^2[\cosh(2\epsilon)+\cos\vartheta],\nonumber\\
\tau(k)={}&1-M_1/r^2,\qquad
\tau(k^2)=1-2M_1/r^2+M_2/r^4.
\label{eq:scalarcoeff}
\end{align}
The $\cos\vartheta$ term already records interference between translation and Fourier steps. Higher orders follow the same finite recurrence in Eq.~\eqref{eq:twist}.

The elementary expression $\log(|a_\alpha|/2)$ benchmarks the projected calculation. Equations~\eqref{eq:mainseries} and~\eqref{eq:extract} evaluate the coupled ladder at finite $\Omega$.

\section{Determinant factorization and error control}
\label{app:walkproof}
Set $A(x)=EI-\Omega K-\lambda\cos(x+i\epsilon)Q$. The recurrence for Eq.~\eqref{eq:Fwalk} has transfer matrix
\begin{equation}
T_h(x)=\begin{pmatrix}e^{-h}A(x)&-e^{-2h}I\\I&0\end{pmatrix}.
\end{equation}
It is conjugate by the constant matrix $\operatorname{diag}(e^{-h}I,I)$ to $e^{-h}T_0(x)$. Its exponents are therefore $\gamma_j-h$ and $-\gamma_j-h$. With $I_6$ the $6\times6$ identity, Schur elimination of the lower-right block $U I$ in $U I_6-T_h$ gives
\begin{align}
\log\Delta(U I_6-T_h)
&=\log\Delta(U-e^{-h}A+e^{-2h}U^{-1})\nonumber\\
&=\mathcal D_{\epsilon,h}-3h.
\end{align}
Here $\Delta(U)=1$, and the triangular elimination factors have determinant one. Deninger's cocycle determinant theorem (Theorem~9 and Eq.~(32) of Ref.~\onlinecite{Deninger2011}) identifies the left side with the sum of the positive parts of the six exponents. Since $h\geq0$, this gives Eq.~\eqref{eq:Dchannels}. Products based on the inverse rotation are related to the usual transfer products by shifting the initial phase before phase averaging. For $\mathbb J=\begin{pmatrix}0&I\\-I&0\end{pmatrix}$, the symmetry $A^T=A$ implies $T_0^T\mathbb J T_0=\mathbb J$ and hence reciprocal singular-value pairs for every transfer product, including at complex phase.

A diagonal similarity removes the bias on a finite open chain, as in the imaginary gauge transformation of Hatano and Nelson~\cite{HatanoNelson1996}. The corresponding infinite-chain transformation is unbounded. The canonical determinant is evaluated by taking local closed moments before the logarithmic limit.

For $\mathscr K$ in Appendix~\ref{app:walkconstruction}, the support of $\mathscr K^m$ lies in $|s|,|k|\leq2m$, containing at most $d_m=(4m+1)^2$ coefficient positions. Parseval's identity and the triangle inequality imply
\begin{equation}
\|\mathscr K^m\|\leq\sqrt{d_mm_{2m}}.
\end{equation}
Consequently $d_mm_{2m}<1$ is a finite sufficient condition for
\begin{equation}
\|\mathscr K\|\leq
(d_mm_{2m})^{1/(2m)}<1,
\end{equation}
which supplies a tail bound in Eq.~\eqref{eq:tailwalk}. At a true real spectral energy with $\epsilon=h=0$, noninvertibility gives $\|\mathscr K\|=1$, placing this case outside the strict-contraction domain of Eq.~\eqref{eq:tailwalk}. An exponent enclosure requires two-sided determinant bounds and channel separation. If certified intervals $[l_0,u_0]$ and $[l_h,u_h]$ enclose the two determinants and Eq.~\eqref{eq:separator} is certified, then
\begin{equation}
l_0-u_h+h\leq\gamma_3\leq u_0-l_h+h.
\end{equation}

For a normalized finite-chain vector $\psi$, let $\widetilde\psi$ be its zero extension, $H_\theta$ the infinite real-phase ladder Hamiltonian, and $\Sigma_\lambda$ its spectrum. Then
\begin{equation}
\operatorname{dist}(E,\Sigma_\lambda)
\leq\|(H_\theta-E)\widetilde\psi\|.
\end{equation}
For identity longitudinal hopping, the squared residual is the sum of the internal eigenvector residual squared and the squared norms of both end-cell vectors. Finite-chain eigenvalues supply the energy samples for the growth-rate calculations.

\section{Participation distributions near a projected edge}
\label{app:distribution}

\begin{figure}[H]
\includegraphics[width=\columnwidth]{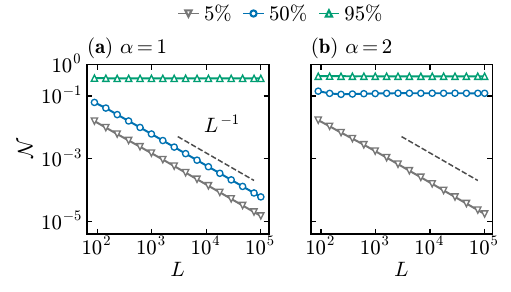}
\caption{Complete-band NPR quantiles at $\lambda=12.55$, $\Omega=100$, and $\eta=1$ for bands (a) 1 and (b) 2. The fifth, fiftieth, and ninety-fifth percentiles pool all states and four phases separately at each of the 16 lengths $L=89$--$100000$. Dashed $L^{-1}$ guides are normalized to 2.5 times the band-1 median at $L=100000$.}
\label{fig:distribution}
\end{figure}

At $\lambda=12.55$, the full reference model ($\Omega=100$, $\eta=1$) has phase-pooled NPR distributions whose quantiles follow different finite-size trends [Fig.~\ref{fig:distribution}]. This point lies just outside the projected band-1 interval and just inside the projected band-2 interval. All states and four phases are pooled and ranked separately at each size. In band 1, increasing $L$ from $1597$ to $100000$ reduces the median from $0.00380747$ to $6.09634\times10^{-5}$, while the ninety-fifth percentile changes from $0.362393$ to $0.362934$. The fraction above $\Ncell=0.1$ changes from $37.8522\%$ to $38.1820\%$. In band 2, the fifth percentile decreases from $0.00108614$ to $1.73772\times10^{-5}$, while the median changes from $0.121295$ to $0.121764$ and the ninety-fifth percentile from $0.423547$ to $0.423517$.

The quantile separation also occurs within individual phases. At $L=100000$, the band-1 fifth percentiles span $1.50767\times10^{-5}$--$1.50774\times10^{-5}$, while the ninety-fifth percentiles span $0.362920$--$0.362939$. The fractions above $\Ncell=0.1$ span $38.1810\%$--$38.1830\%$ in band 1 and $52.1440\%$--$52.1520\%$ in band 2.

Table~\ref{tab:edgefits} gives the exponents at this point for three fitting ranges.

\begin{table}[H]
\caption{Band-averaged exponents at $\lambda=12.55$ for the full reference model. The fits use four, four, and three lengths in the indicated ranges, respectively; the largest length is $100000$, and the other lengths are Fibonacci numbers. Brackets span the four separate phase fits for the same lengths, with endpoints rounded outward.}
\label{tab:edgefits}
\begin{ruledtabular}
\normalsize
\begin{tabular}{cccc}
Fitted $L$ & $\alpha$ & $D_{2,\alpha}$ & Phase range\\
\hline
$17711$--$75025$ & $1$ & $0.383139$ & $[0.382873,0.383506]$ \\
$17711$--$75025$ & $2$ & $0.530813$ & $[0.530332,0.531158]$ \\
$28657$--$100000$ & $1$ & $0.382829$ & $[0.382527,0.383232]$ \\
$28657$--$100000$ & $2$ & $0.530048$ & $[0.529714,0.530320]$ \\
$46368$--$100000$ & $1$ & $0.382719$ & $[0.382258,0.383288]$ \\
$46368$--$100000$ & $2$ & $0.529748$ & $[0.529282,0.530229]$ \\

\end{tabular}
\end{ruledtabular}
\end{table}

The exact finite-ensemble relation
\begin{equation}
\log\Icell_{\mathrm{geom},\alpha}(L)=-\log L-\langle\log\Ncell\rangle_{\alpha,L}
\end{equation}
connects this distribution to the band-averaged exponent, with the same band and phase average on both sides. As an illustrative asymptotic mixture, a size-independent fraction $f$ with IPR proportional to $L^{-1}$ and a fraction $1-f$ with saturating IPR give $D_{2,\alpha}=f$. Here $f$ denotes an asymptotic scaling fraction. The band-averaged exponent characterizes the mixture, while the threshold fractions above describe finite-$L$ NPR.

\section{Numerical protocols and convergence}
\label{app:numerical}
\subsection{Complete spectra, phases, and energy samples}
\label{app:staticdetails}
The Hamiltonian is real symmetric with half-bandwidth three. Complete banded eigenvalues and inverse iteration with reorthogonalization and Rayleigh residual refinement give every eigenstate, including boundary states, for the complete-band participation statistics. Fixed-size main results use $L=1597$, and $L=987$ is a separate size check. The six representative parameters and $\lambda=12.55$ use 16 lengths from $L=89$ to $100000$, with primary fits on $L=28657,46368,75025,100000$. The 281-point exponent scan is fitted to $L=10946,17711,28657,46368$. Static phases are $\theta_\ell=0.2718+2\pi\ell/M_\theta$ with $M_\theta=4$ and the left-cell origin. The same numerical value of $(\sqrt5-1)/2$ is used at every chain length, without replacing it by a size-dependent rational approximant. Band separation was checked before assigning $L$ energy-ordered states to each band.

At $L=100000$, the arithmetic mean cell NPR in each lower band, over all states and four phases, ranges from $0.494296$ to $0.621515$ at $\lambda=1,4.7,12$, and from $1.22448\times10^{-5}$ to $3.05133\times10^{-5}$ at $\lambda=3.1,8,14$.

For Fig.~\ref{fig:paths}, complete banded spectra check band separation, and $LDL^T$ inverse iteration with local reorthogonalization reconstructs every state in both lower bands, with full-Hamiltonian residuals below $10^{-9}$. Independent pivoted banded solves reproduce every $\Ncell>0.1$ classification in 31 individual-phase checks at $L=1597$, covering the four passing endpoints of all seven paths and three additional controls. Small-system direct diagonalization provides a separate solver check.

All seven paths use the full coupled Hamiltonian, four phases, and every eigenstate in both lower bands at $L=987,1597$. The path values are computed on a separate mesh, without interpolation from the heat maps. The paths use $\Delta u=0.025$ near each window, with threshold brackets refined to $0.005$; low-modulation and inter-window localized controls are also included. The offset controls are evaluated at $u\geq1$, so their modulation amplitudes remain positive.

The maps use $L=1597$ and the same four phases, with a uniform $\Delta u=0.05$, $\Delta v=0.125$ mesh. Neighborhoods whose sampled $f_{\min}$ range exceeds $0.08$, or whose central fraction lies between $0.05$ and $0.95$, are refined to spacings $0.025$ and $0.0625$. Each displayed rectangle takes the value of its actual sample; neighboring values are not interpolated. The white curves $|v q_\alpha(u)|=2$ are isolated-channel benchmarks for comparison with the sampled full-model interiors.

\begin{table}[H]
\caption{Consecutive sampled $u$ intervals satisfying the 90\% criterion in both lower bands at both $L=987,1597$. The adjacent failed samples are $0.005$ below each left endpoint and $0.005$ above each right endpoint. These are finite-sample threshold brackets. The paths $v_\pm$ are defined in Eq.~\eqref{eq:curvedpath}; full sampling counts and endpoint records accompany the data.}
\label{tab:paths}
\begin{ruledtabular}
\normalsize
\begin{tabular}{ccc}
Path & Window 1 & Window 2\\\hline
$v=u$ & $4.180$--$5.070$ & $11.720$--$12.365$ \\
$v=0.95u$ & $4.125$--$5.100$ & $11.685$--$12.395$ \\
$v=1.05u$ & $4.225$--$5.045$ & $11.755$--$12.335$ \\
$v=u-0.25$ & $4.110$--$5.095$ & $11.705$--$12.375$ \\
$v=u+0.25$ & $4.235$--$5.045$ & $11.735$--$12.350$ \\
$v_-(u)$ & $4.150$--$5.065$ & $11.730$--$12.375$ \\
$v_+(u)$ & $4.215$--$5.075$ & $11.710$--$12.355$ \\

\end{tabular}
\end{ruledtabular}
\end{table}

At $L=1597$, increasing the static phase count from four to eight at 11 reference samples changes the lower-band fraction above $\Ncell=0.1$ by at most $0.00226988$; the four refined reference endpoints still pass the $90\%$ criterion. The reported two-length interiors use the specified four-phase ensemble.

Each QR/Gaussian pair uses the same full-precision energy of the 799th state within a band of an open chain with $L_E=1597$ and $\theta=0.2718$; ``band-middle'' denotes spectral rank. The 21 reference samples combine $\lambda=1,3.1,4.7,8,12,14,12.55$ with all three bands; 18 additional second-band energies sample $(\Omega,\eta)$ controls. The separate molecular-separation scan contains 150 second-band energies, for two $\lambda$ values and 75 values of $\Omega$. All use $\epsilon=0$ and $h=2.5$. Over both retained transfer lengths and starting phases in the 150-sample set, the smallest finite-QR separation margins are $h-\widehat\gamma_3=2.37699$ and $\widehat\gamma_2-h=1.31105$. For identity longitudinal hopping, zero extension of a normalized finite-chain vector $\psi$ gives
\begin{equation}
r_{\rm ext}^2=\|(H_L-E)\psi\|^2+\|\psi_0\|^2+\|\psi_{L-1}\|^2.
\end{equation}
Here $H_L$ is the finite-chain Hamiltonian, and $\psi_0$ and $\psi_{L-1}$ are its end-cell amplitude vectors.
Across the 21 reference and 18 coupling-diagnostic energies, the largest $r_{\rm ext}$ is $0.0496733$. Including the endpoint amplitudes, this residual bounds the distance to the infinite spectrum [Appendix~\ref{app:walkproof}]; it differs from the finite-matrix eigenvector residual.

\subsection{Moments and transfer convergence}
\label{app:walknumerics}
Uniform phase sampling is the periodic trapezoidal rule~\cite{TrefethenWeideman2014}. The Gaussian phase mesh is $\theta_\ell=0.2718+2\pi\ell/M_\theta$, referred to the central cell of the finite moment support. The Lanczos starting vector lies in the direct sum of these phase spaces. Each starting orbital defines a unit-mass measure; their sum gives $\tau(I)=3$. One Gaussian rule is constructed for each phase-averaged orbital measure. Phase averaging precedes quadrature construction; reversing these operations gives a different finite-order estimator. A support of $4n+7$ cells contains all paths required through moment degree $2n-1$. A uniform phase mesh with $M_\theta>4n-2$ integrates these finite moments exactly in exact arithmetic; for $n=512$ the sufficient count is 2048, whereas $n=1024$ would require more than 4094 phases. The 1024-node, 512-phase refinement uses finite phase quadrature.

Low-order moments were independently evaluated by noncommutative coefficient multiplication at the original irrational frequency. The power series uses powers of $I-\mathsf J_n$, where $\mathsf J_n$ is the $n$-dimensional Lanczos Jacobi matrix, avoiding subtraction of large alternating raw moments. Gaussian rules retain unmodified positive nodes, with local reorthogonalization and stronger small-order checks. The unbiased/biased rules use 512/128 nodes and 512 phases. Signed finite-length QR estimates are retained; the same-energy difference $\Delta_3=\widehat\gamma_3^{\rm Gauss}-\widehat\gamma_3^{\rm QR}$ can have either sign. Transfer-length, node, and phase refinements quantify separate numerical sensitivities. Rigorous enclosures require the determinant and separation bounds in Appendix~\ref{app:walkproof}.
In the 21-sample reference set, the largest $|\Delta_3|$ is $0.00307085$ at $\lambda=12.55$ in band 2. Across all 39 reference and coupling-diagnostic energies, the finite-QR bias margins satisfy $h-\widehat\gamma_3\geq0.563256$ and $\widehat\gamma_2-h\geq0.880809$, supporting channel selection at these sampled energies.

Doubling the retained QR length from $80000$ to $160000$ cells changes $\widehat\gamma_3$ by at most $3.57578\times10^{-5}$. The largest two-phase spread at the longer length is $3.39318\times10^{-5}$; changing the transient from 300 to 1000 cells changes the estimate by at most $3.97958\times10^{-5}$. Table~\ref{tab:gaussconv} separates node and phase sensitivity of the unbiased determinant at fixed energy.

\begin{table}[H]
\caption{Unbiased determinant refinements for $\Omega=100$ and $\eta=1$, at fixed energy within each row. The baseline is $\mathcal G_{512,512}$; $\delta_{n,M_\theta}=\mathcal G_{n,M_\theta}-\mathcal G_{512,512}$ resolves node and phase sensitivity. The two subscripts give node and phase counts.}
\label{tab:gaussconv}
\begin{ruledtabular}
\normalsize
\setlength{\tabcolsep}{2pt}
\begin{tabular}{cccc}
$\lambda,\alpha$ & $\delta_{512,256}$ & $\delta_{512,2048}$ & $\delta_{1024,512}$\\\hline
$4.7,1$ & $-1.71241{\cdot}10^{-12}$ & $6.71463{\cdot}10^{-13}$ & $-4.04701{\cdot}10^{-4}$ \\
$12,1$ & $-4.63629{\cdot}10^{-13}$ & $1.79234{\cdot}10^{-12}$ & $-5.10589{\cdot}10^{-4}$ \\
$12.55,2$ & $-9.71079{\cdot}10^{-6}$ & $6.78720{\cdot}10^{-8}$ & $-1.89115{\cdot}10^{-3}$ \\
\end{tabular}\end{ruledtabular}\end{table}

The molecular-separation maps use 75 values of $\Omega$ from $15$ to $200$ in steps of $2.5$ and 41 values of $\eta$ from zero to one, at $u=v=12,12.55$. Each point contains every second-band state at $L=1597$ and four phases; the $\eta=1$ cuts also use $L=987$. At $\lambda=12.55$, $\Omega=50$, the median NPR decreases from $0.0239478$ to $0.0160068$ between these lengths.

QR products retain $160000$ cells after a 300-cell transient and average the fixed starting phases $\theta=0.173,0.731$. Across the 150 band-middle samples, doubling the retained length from $80000$ cells changes the phase-averaged minimum by at most $4.00527\times10^{-5}$; the largest two-phase spread is $3.17565\times10^{-5}$. The maximum absolute same-energy Gaussian--QR discrepancy is $0.00857001$ at $\lambda=12.55$, $\Omega=15$. At eight selected static samples, increasing the phase count from four to eight changes the mean NPR by at most $0.00103891$.

At the weak-attenuation sample $\lambda=12.55$, $\Omega=50$, increasing the unbiased rule from 512 to 1024 nodes at 512 phases changes the extracted minimum from $0.0107164$ to $0.00693156$, compared with QR $0.00402936$. Increasing the phase count alone to 2048 changes the result by $8.79689\times10^{-8}$. The biased rule stays at 128 nodes and 512 phases; node truncation remains material at this sample.

At the 150 band-middle energies in Fig.~\ref{fig:interband_response}(d), series extraction uses unbiased/biased degrees $1023/255$, matched to the moment degrees of the $512/128$-node Gaussian rules, with 512 phases. For canonical moments and under Eq.~\eqref{eq:separator}, the determinant truncation errors $e_h=\mathcal S_{N_h}(E,h)-\mathcal D_{0,h}(E)$ give
\begin{equation}
\widehat\gamma_3^{\rm Series}-\gamma_3=e_0-e_{2.5}.
\label{eq:series_extraction_error}
\end{equation}
Nonnegative determinant errors can therefore produce a signed extraction error; cancellation between them can bring the extracted value close to QR before either determinant has converged.

Near zero singular values, the factors $(1-x)^k$ make the logarithmic series converge slowly. At matched moment degree, Gaussian deviations of the unbiased determinant from the finite-length QR reference are smaller than series deviations at every sampled $\Omega$ for both $\lambda$ values. This ordering concerns a single determinant; their difference retains the separate node and phase sensitivities above. The real-spectrum unbiased determinant has $\|\mathscr K\|=1$, outside the strict-contraction domain of the finite tail bound in Eq.~\eqref{eq:tailwalk}. Gaussian convergence for canonical moments is given by Eq.~\eqref{eq:Gaussbound}.

\subsection{Long-time dynamics and the order of limits}
Dynamical phases are $\theta_\ell=0.2718+2\pi\ell/M_\theta$ relative to the launch cell, with $M_\theta=32$. The 16-phase check selects alternate phases, and the 64-phase check adds midpoints. The launch density matrix is $\rho_0=\frac13\sum_s|m_0,s\rangle\langle m_0,s|$, evolved through a complete spectral expansion over all three bands. Independent stratified meshes of 128 and 256 points estimate late averages over $5\times10^4\leq Jt\leq10^5$. The late RMS is $[\overline{\langle M_2\rangle}]^{1/2}$, with time averaging before the square root; retention is $\overline{\langle P(100,t)\rangle}$.

For a nondegenerate finite chain, the diagonal ensemble gives
\begin{equation}
\overline p_m^{(L)}=\frac13\sum_n p_{m_0}^{(n)}p_m^{(n)}.
\end{equation}
It gives the infinite-time average at fixed $L$, providing a separate reference for the finite-window data. The thermodynamic limit requires an additional limit in $L$. For degenerate levels, coherence within each energy subspace must be retained. The computed level spacings are nonzero at the working numerical precision and exceed the residual-based eigenenergy scales for the corresponding samples.

The probability in the first and last ten cells is $B_{10}$. The common early interval $Jt\leq230$ satisfies $B_{10}<10^{-8}$ for every compared trajectory, a two-length retention difference below $10^{-3}$, and a relative RMS difference below $10^{-3}$ (absolute tolerance $10^{-8}$ cells for very small RMS). The next sampled time violates the common criteria; later evolution describes the finite chain. Normalization, energy conservation, interval monotonicity, and the complementary-probability identity $Q_{20}(t)=1-P(40,t)$ were checked over the full time range; $P(100,t)$ was evaluated separately.
Across the trajectories, the maximum normalization error is $3.04268\times10^{-12}$ and the maximum energy-conservation discrepancy is $2.52749\times10^{-9}J$. Independent direct diagonalization and small-chain matrix-exponential evolution check the propagated observables; the residual-based eigenenergy scale multiplied by $t_{\max}$ is at most $1.44315\times10^{-5}$.

For the 9 parameters at $L=1597$ in Fig.~\ref{fig:dyn}, independent 128/256-point late-time meshes change retention by at most $0.000537864$ and RMS by $0.307381$ cells. Nested 16/32-phase ensembles on the 256-point mesh differ by at most $0.00528931$ in retention and $0.725334$ cells in RMS. The largest retention difference between the 256-point time average and the fixed-chain diagonal ensemble is $0.00161586$.

Relative phase changes use $|\sigma_{2M_\theta}-\sigma_{M_\theta}|/\sigma_{2M_\theta}$. Among the localized controls, the 16-to-32-phase RMS change reaches $17.2873\%$ at $\lambda=8$; the 32-to-64-phase check still changes the RMS by $20.0945\%$ at $\lambda=3.1$. The reported widths describe the specified 32-phase ensemble, with sub-cell or few-cell confinement at the localized controls and spreading over hundreds of cells at the reentrant samples.

\subsection{Band-filtered launch checks}
\label{app:banddyn}
The filtered calculation in Fig.~\ref{fig:banddyn} uses $\lambda=4.7,12$, both lengths, and the same 32 centered phases as Fig.~\ref{fig:dyn}. For each band and phase, the three unnormalized vectors $\Pi_\alpha|m_0,s\rangle$ are evolved independently. Their cell probabilities are summed and divided by $\sum_s\|\Pi_\alpha|m_0,s\rangle\|^2$, implementing Eq.~\eqref{eq:bandlaunch}. Normalizing the three projected vectors separately before taking an equal average would change the conditional ensemble.

Independent dense diagonalization and direct matrix-exponential evolution at $L=21$ check the filtered observables. Every phase-resolved conditional ensemble satisfies $B_{10}<10^{-8}$, and the two-length RMS comparison meets the common early-time criterion through $Jt=230$. The next sampled time fails the common endpoint threshold; later curves include boundary effects.

The filtered late averages use a separate 128-point stratified mesh over $5\times10^4\leq Jt\leq10^5$, compared with a nested 64-point subset and the diagonal ensemble. At $L=1597$, the 16-to-32-phase third-band RMS changes are approximately $15\%$ and $13\%$ at $\lambda=4.7$ and $12$, respectively. These widths describe the filtered 32-phase ensemble and remain on a much smaller spatial scale than the lower-band spreading.

\subsection{Matched-criterion endpoint shifts}
\label{app:matchedendpoints}
The $\eta=0$ and $1$ comparisons retain the open boundaries, irrational frequency, left-cell phase origin, and four static phases of Appendix~\ref{app:staticdetails}. At $\Omega=100$, both lower bands are tested separately at $L=987,1597$. The joint passing fraction is
\begin{equation}
F_\eta(\lambda)=\min_{\substack{\alpha=1,2\\L=987,1597}}
\frac{1}{4L}\sum_{\ell=0}^{3}\sum_{n\in\alpha}
\mathbf1[\Ncell_{n,\ell}^{(\eta)}(\lambda)>0.1],
\label{eq:matchedfraction}
\end{equation}
with a passing control satisfying $F_\eta\geq0.9$. The four phase-pooled fractions enter this minimum without averaging across bands or lengths.

At $\eta=0$, each channel is solved as a tridiagonal chain. A normalized channel wave function $\phi_m$ has cell NPR $[L\sum_m|\phi_m|^4]^{-1}$; the uniform molecular energy shifts the eigenvalues. Spectral clusters remain separated. The four phases form a fixed deterministic ensemble: $\theta$ and $\theta+\pi$ give identical complete-band NPR distributions in a decoupled channel.

A common mesh covers $\lambda\in[3.8,5.5]$ and $[11.2,12.9]$ with spacing $0.025$. All observed pass/fail transitions of either $\eta$ are refined on the same $0.005$ mesh, giving 170 controls for each model and length. Each window is a single consecutive sampled passing block containing the shared controls $\lambda=4.7$ and $12$. Table~\ref{tab:matchedendpoints} propagates the adjacent failed/passed endpoint brackets by interval subtraction. The resulting ranges estimate the grid resolution of the sampled NPR transitions.

Among the 16 adjacent failed/passed boundary controls, the smallest passing $F_\eta$ is approximately $0.907295$ and the largest failing $F_\eta$ is approximately $0.885258$. Independent recalculations of all 32 boundary samples, retaining all four phases and both complete lower bands, change no NPR-threshold classifications. A separate tridiagonal eigensolver checks the decoupled channels, while pivoted banded inverse iteration checks the full-model states. The largest NPR difference is $7.54687\times10^{-11}$, compared with a minimum single-state distance of $3.40728\times10^{-6}$ from $\Ncell=0.1$. Counts use unrounded values.

\bibliography{trimer_paper}
\end{document}